   \documentclass[12pt]{article}
\usepackage{float}
   \usepackage[round,authoryear]{natbib}

 \usepackage{soul}   

 \usepackage{hyperref}

  \usepackage[latin1]{inputenc}
 \usepackage{amsfonts}
 \usepackage{amsthm}
 \usepackage{amssymb, latexsym, amsmath}
 \usepackage{amsbsy}
\usepackage{amstext}
\usepackage{amsopn}
\usepackage{amsgen}
 \usepackage{graphicx}
 \usepackage[normalem]{ulem}
   \usepackage{wasysym}
   \usepackage{deluxetable}
  \usepackage{verbatim}
  \usepackage[T1]{fontenc}
  \usepackage[bottom]{footmisc}

  \usepackage{subfigure}

   \usepackage[dvips]{color}
   \usepackage{color}

\definecolor{mypink}{rgb}{0.858, 0.188, 0.478}
\definecolor{mygrey}{rgb}{0.55, 0.68, 0.55}

 \newcommand{\s}{\nobreak\hspace{.11em}\nobreak}

 \newcommand{\be}{\begin{equation}}
 \newcommand{\ee}{\end{equation}}
 \newcommand{\ba}{\begin{eqnarray}}
 \newcommand{\ea}{\end{eqnarray}}
 \newcommand{\bs}{\begin{subequations}}
 \newcommand{\es}{\end{subequations}}

     \newcommand{\Omegabold}{\boldsymbol\Omega}

  \newcommand{\eRbold}{\mbox{{\boldmath $
  {R}$}}}

\begin{document}

  \title{
          {\Large{\textbf{{{A synchronous moon as a
          possible cause\\ of Mars' initial triaxiality
          }
 ~\\
 ~\\}
            }}}}
 \author{
                                          {\large{Michael Efroimsky}}\\
                                          {\small{US Naval Observatory, Washington DC 20392}}\\
                                          {\small{michael.efroimsky$\s$@$\s$gmail.com}}
 }
     \date{}

 \maketitle

 \begin{abstract}

\noindent
 The paper addresses the possibility of a young Mars having had a massive moon, which synchronised the rotation of Mars, and gave Mars an initial asymmetric triaxiality to be later boosted by geological processes. It turns out that a moon of less than a third of the lunar mass was capable of producing a sufficient initial triaxiality. The asymmetry of the initial tidal shape of the equator depends on timing: the initial asymmetry is much stronger if the synchronous moon shows up already at the magma-ocean stage. From the moment of synchronisation of Mars' rotation with the moon's orbital motion, and until the moon was eliminated (as one possibility, by an impact in the beginning of the LHB), the moon was sustaining an early value of Mars' rotation rate.

 \end{abstract}

 \section{Motivation.\\ The equatorial ellipticity of Mars\label{Hypothesis}}

 Mars' triaxiality makes itself most evident through the equatorial ellipticity produced by the Tharsis Rise and by a less prominent elevation located almost diametrically opposite to Tharsis and constituted by Syrtis Major Planum and an adjacent part of Terra Sabaea \citep{smith} .

 Terra Sabatea being more ancient, Syrtis Major's present form is dated by crater counts to the early Hesperian epoch \citep{syrtis}, and is a smoothly sloping shield-volcano dome of an average $2.1$~km height, its peak extending to 2.3 km.

 The largest highlands province on Mars and in the solar system, Tharsis is rising, in average, to about $7$~km, excluding the much higher volcanoes.
 Its formation had begun at least in the Noachian period, and continued through the entire Hesperian, the volcanoes staying active into the Amazonian epoch. As was pointed out by \citet{zuber}, this rise may mask or modify gravitational signatures that contain important information on the Martian
 gravitational field as a whole. For this reason, those authors produced a ``Mars without Tharsis'' gravitational field, by subtracting the gravitational signature of Tharsis from the full field of the planet. Even then, Mars retained much of its triaxiality.
 \vspace{1.5mm}

\begin{figure}[H]
  \includegraphics[width=1.21\textwidth]{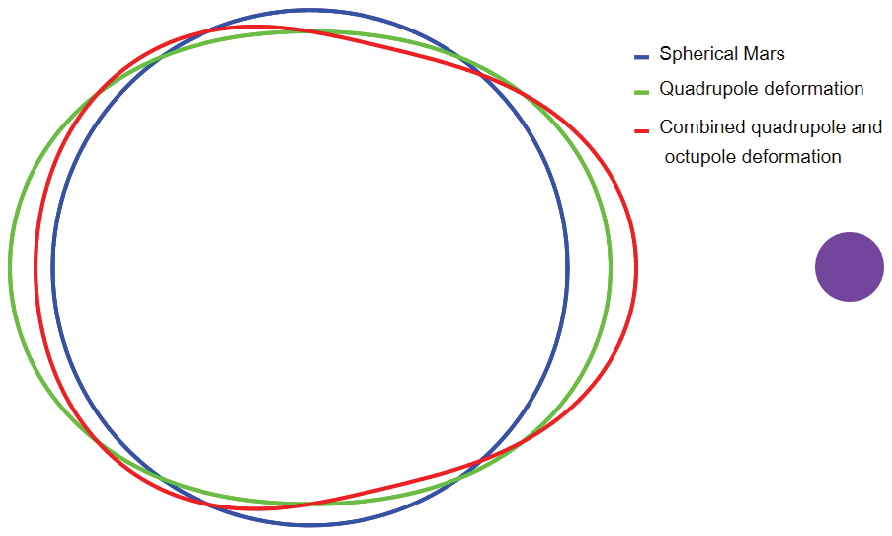}
             \caption{\small Mars deformed by its putative moon~Nerio.~The~blue~line~depicts a~spherical~Mars.~In~an~exaggerated way,~the~green~line~shows~a~symmetric,~quadrupole deformation. The red line shows a combined quadrupole~and~octupole~distortion.}
         \label{figure}
     \end{figure}

 \noindent
 To explain the origin of this shape, we propose a hypothesis consisting of two items:
 \begin{itemize}
 \item[(a)]  An initial, `seed' triaxiality was created by a massive moon orbiting a young and still plastic Mars on a synchronous orbit.
             Showing the same face to the moon, Mars assumed a shape close but not identical to a triaxial ellipsoid, its longest axis aligned with the moon. While a symmetrical ellipsoidal part of the shape was created by the quadrupole component of the tide, a weaker,
             antisymmetrical perturbation was added to the shape by the octupole component, see Figure \ref{figure}.
 \item[(b)]  After the moon produced the seed triaxiality and asymmetry of Mars, the tidally elevated provinces became, more than others, prone to
             convection-generated uplifts and tectonic and volcanic activity. These processes began to gradually add to the equatorial ellipticity. Owing to the degree-3 initial asymmetry of the shape, they were not acting in a symmetrical way; hence the resulting height disparity between Tharsis and Syrtis Major.
 \end{itemize}

 \noindent
  In this paper, we address item (a) and demonstrate that a synchronous moon of less than a third of the lunar mass was capable of providing a seed triaxiality.
  The seed asymmetry of the equator was considerable if the synchronous moon existed already at the magma-ocean epoch, and was weaker if the moon showed up at the solidification stage.

  Checking on item (b) requires quite a programme of research, but a qualitative argument in support of this item is available readily for a moon
  whose orbit is synchronous but not circular.
  A finite eccentricity causes the tidal deformation to oscillate and produce heat, because an elliptic synchronous orbit makes the moon's position appear to oscillate E-W in the sky over the same region (or to describe an analemma if the inclination is nonzero).
  The submoon and antimoon zones might have been warmer than the rest of the surface~---~though this outcome is very structure-dependent and is not warranted \citep{segatz88}.
  Modeling of a young Kelvin-Voigt Moon has demonstrated that the closer the Moon was to the Earth the higher the ratio of near to farside tidal heat flux, the difference being in the dozens of percent \citep[Figure 6]{Quillen}.  This motivates us to extend this observation to Mars, and to expect that the degree-3 tide made the submoon zone warmer than the antimoon zone, with the possibility of a plume emerging in the submoon zone, as Tharsis' predecessor.  It should be added, though, that the cited model contains also an argument {\it against} our hypothesis, an increase of the heat flux toward the polar zones~---~which poses a question why the volcanic activity near the poles was lower than on Tharsis. To answer this question, accurate modeling based on a more realistic rheology is required.


  As per advice from Beno\^{i}t Noyelles, who kindly reviewed this paper and contributed greatly to its improvement, we christen the hypothetical moon \textit{Nerio},
 after a war goddess who was Mars' partner in ancient cult practices, later to be supplanted by deities adapted from other religions.

 \section{Whence had it come, whither gone?}

 \subsection{Origin}

  Aside from the obvious possibility of {\it in situ} formation, the putative moon may as well had been captured in the remnants of the protoplanetary disk (or in a later debris disk,
  if it existed) and was then slowed down by friction, and eventually synchronised by tides. For example, \citet{Hunten}  demonstrated that while slowly-rotating condensation in the solar nebula is unlikely to capture a moon, the odds are greatly improved if the atmosphere is rapidly rotating.  However, given the relatively short life spans of these disks, more probable option is a megaimpact. Had the impact happened during the magma-ocean stage, it would hardly have influenced the subsequent development of Mars' global structure. On the other hand, had it happened during the formation of crust, it may have, speculatively, left some signature~---~whence the question arises whether that impact could be the one responsible for the north-south hemispherical dichotomy, a theme beyond the scope of our study.

  As we shall see shortly, a moon of less than a third of the lunar mass was sufficient to produce the required initial triaxiality.
  Creation of such a moon by a collision of an impactor with an Earth-sized planet would not be a problem.
  Mars, however, is more than nine times less massive than the Earth, which reduces the probability of generation of the needed moon via impact.
  This obstacle can be sidestepped. As was shown by \citet{rufu}, our Moon could have been created by a succession of smaller collisions.  Each such collision resulted in a debris disk accreting to produce a moonlet.  The moonlets tidally receded outward and coalesced into the Moon. This mechanism should be applicable to any terrestrial planet, including Mars.

  One more mechanism of acquisition of a massive satellite is binary-exchange capture. Originally developed for giant planets \citep{Agnor}, it was recently applied
  to terrestrial planets. It turned out that smaller planets capture satellites more efficiently because of the slower encounter velocities
  in their weaker gravity wells \citep{Williams_and_Zugger}.

  %

  Whatever its origin, we assume that the moon Nerio was synchronised sufficiently early, when Mars was not completely rigid~---~that is to say, either during the magma-ocean stage or during the subsequent period of solidification while the lithosphere was still weak \citep{matsuyama}.  Also, it was not necessary for the moon to have been born or captured exactly at the then synchronous radius. Sufficient would have been its emergence in some vicinity of the synchronous orbit~---~whereafter the tidal forces would have done the rest. For a molten or semimolten planet, the value of $k_2/Q$ could have been of order $0.03$ to $0.6$. [\,See equation (\ref{k2}) in Appendix \ref{A2}, and mind that the $Q$ of such a planet can reside anywhere between $\simeq 2$ to $\simeq 10\s$. For an extended explanation, see Appendix \ref{deta}.\s] So tidal dissipation in the planet was two to three orders of magnitude higher than presently. Had Nerio begun its life slightly above (below) synchronism, it would have quickly synchronised Mars' rotation while tidally receding (descending). Both options are described by the same equation following from the angular momentum conservation law \citep{pathways}.

 \subsection{Disposal}

  As one possibility, Nerio could have been destroyed by the Late Heavy Bombardment (LHB)~---~in which case Phobos and Deimos may be this moon's remnants or, more likely, the remnants of a larger fragment eliminated by a later event \citep{phobosdeimos}.

  A potential restriction on the timing of LHB-caused disposal is that the destruction event ought to have happened early in the LHB, to enable the subsequent LHB to smoothen the latitudinal dependence of the resulting crater distribution. The ancient crater population seen in the southern highlands and in the Quasi-Circular Depressions in the north (buried impact basins) shows no such latitudinal dependence. So the downfall of Nerio's remnants should have predated those provinces' formation which coincided in time with the LHB. We are grateful to Andrew Dombard for drawing our attention to this circumstance.


  \section{Formulae\label{Formulae}}

 Consider a static configuration comprising two mutually synchronised bodies: a planet of mass $M$ and radius $R\s$, and a moon of mass $M^{\,\prime}$.  Aiming at crude estimates, we set the eccentricity zero, so the semimajor axis coincides with the distance $r$ between the centres of mass of the partners. This enables us to employ a simple formalism developed for static tides with no lag.

 \begin{table*}
 \caption{~~~~Symbol key}
 \label{description}
   \hskip-3.2cm
 \begin{tabular}{@{}llll@{}}
 \hline\\
   \vspace{2mm}
 Variable & ~~~~~Value & Explanation & Reference \\
 \hline \\

 \vspace{1.4mm}

 $G$ & $6.67430 \times 10^{-11}$ $\frac{\mbox{m}^{\s 3}}{\mbox{kg~\s s}^{\s 2}}$ & gravitational constant & \citet{gravitational_constant} \\[2pt]

 \vspace{1.4mm}

 $  M                $   &                                   &  early Mars' mass          &                        \\[2pt]

 \vspace{1.4mm}

 $  R                $   &                                   &  early Mars' mean radius   &                      \\[2pt]

 \vspace{1.4mm}

 $  M^{\,(present)}  $   &  $ 6.4169  \times 10^{ 23} $ kg  &  present Mars' mass        &  \citet{Konopliv2016}  \\[2pt]

 \vspace{1.4mm}

 $  R^{\,(present)}  $   &  $ 3.3895 \times 10^{  6} $  m    &  present Mars' mean radius & \citet{Seidelmann}   \\[2pt]

 \vspace{1.4mm}

 $  M^{\,\prime}  $      &                                   &  Nerio's mass              &         \\[2pt]

 \vspace{1.4mm}

 $  R^{\,\prime}  $      &                                   &  Nerio's mean radius       &       \\[2pt]

 \vspace{1.4mm}

 $ k_2,\s h_2,\s k_3,\s h_3 $  &                            & early Mars' Love numbers          &     \\[2pt]

 \vspace{1.4mm}

 $  Q                       $   &                            & early Mars'  quality factor &          \\[2pt]

 \vspace{1.4mm}

 $  K_2\s \equiv\;\frac{\textstyle k_2}{\textstyle Q} $  &    & early Mars'  quality function &                    \\[2pt]

\vspace{1.4mm}

 $  k_2^{\,\prime}           $   &                            & Nerio's  Love number          &     \\[2pt]

 \vspace{1.4mm}

 $  Q^{\,\prime}             $   &                            & Nerio's  quality factor &          \\[2pt]

 \vspace{1.4mm}

 $  K_2^{\,\prime}\s \equiv\;\frac{\textstyle k_2^{\,\prime}}{\textstyle Q^{\,\prime}} $  &    & Nerio's quality function &                    \\[2pt]

 \vspace{1.4mm}

 $ r $   &     & early Mars' synchronous radius     &     Equation (\ref{r})           \\[2pt]

 \vspace{1.4mm}

 $ r^{\,\prime}_H $   &     & early Mars' reduced Hill radius     &   \citet{pathways}       \\[2pt]

 \vspace{1.4mm}

 $ r_R $   &     & early Mars' Roche radius     &     Equation (\ref{roche})           \\[2pt]

 \vspace{1.4mm}

 $  \dot{\theta} $  &    & Mars' rotation rate &                    \\[2pt]

 \vspace{1.4mm}

  $ \Omega $  &    & the synchronous value of $\dot{\theta}$ &                    \\[2pt]

 \vspace{1.4mm}

 $  \dot{\theta}^{\,\prime}           $   &                            & Nerio's rotation rate          &     \\[2pt]

 \vspace{1.4mm}

 $ a $   &     & Nerio's semimajor axis     &                  \\[2pt]

 \vspace{1.4mm}

 $ n $   &     & Nerio's mean motion     &                  \\[2pt]

 \vspace{1.4mm}

 $ A<B<C $   &     & early Mars' moments of inertia     &                  \\[2pt]

 \vspace{1.4mm}

 $ J_2 $   &     & early Mars'  oblateness     &      Equation (\ref{J2})            \\[2pt]

  \vspace{1.4mm}

 $ J_2^{(present)}  $   &  $1.9566 \times 10^{-3}$   & present Mars'  oblateness     &    \citet{Konopliv_2011}             \\[2pt]

 \vspace{1.4mm}

 $ J_{22} $   &     & early Mars'  triaxiality     &      Equation (\ref{J22})            \\[2pt]

  \vspace{1.4mm}

 $ J_{22}^{(present)} $   &  $6.3106 \times 10^{-5} $   & present Mars'  triaxiality     &     \citet[Tab S4]{Konopliv}             \\[2pt]

  \vspace{1.4mm}

 $ H_2 $   &     & quadrupole tidal elevation     &                  \\[2pt]

   \vspace{1.4mm}

 $ H_3 $   &     & octupole tidal elevation     &                  \\[2pt]

 \hline
 \end{tabular}
 ~\\ \vspace{2mm}
  \end{table*}

 Let $\bf r$ and $\bf R$ be planetocentric vectors pointing, correspondingly, at the moon and at some surface point of the planet.
 In this surface point, the moon-generated perturbing potential is
 \ba
 W({\bf R},\,{\bf r})\,=\,\sum_{l=2}^{\infty}W_l({\bf R},\,{\bf r})\,=\;-\;\frac{G\,M^{\,\prime}}{r}\,\sum_{l=2}^{\infty}\left( \frac{R}{r} \right)^l P_l(\cos\gamma)\,\;,
 \label{W}
 \ea
 $\gamma$ being the angular separation between $\bf R$ and $\bf r$, and $P_l(\cos\gamma)$ being the Legendre polynomials.
 Specifically,
 \ba
 P_2(\cos\gamma)\s=\s\frac{\textstyle 1}{\textstyle 2}\s(3\s\cos^2\gamma - 1) \,\;,
 \quad
 P_3(\cos\gamma)\s=\s\frac{\textstyle 1}{\textstyle 2}\s(5\s\cos^3\gamma - 3\s\cos\gamma)\,\;.
 \label{}
 \ea
 %
 In a static setting, the tidal elevation in the surface point $\bf R$ reads:
 \ba
 H({\bf R},\,{\bf r})\s=\s\sum_{l=2}^{\infty} H_l({\bf R},\,{\bf r})\s=\;\;-\;\frac{1}{g}\;\sum_{l=2}^{\infty} h_l\; W_l(\cos\gamma)\,\;,
 \label{}
 \ea
 where $h_l$ are static quadrupole Love numbers, while the surface gravity is
 \ba
 g\,=\,\frac{G\s M}{R^{\s 2}}\;\;.
 \label{g}
 \ea
 In the submoon point, we have
 $\gamma=0$ and therefore $P_2(\cos\gamma)=P_3(\cos\gamma)=1$.
 In the antimoon one, we get $\gamma=\pi$, whence $P_2(\cos\gamma)= 1$ and $P_3(\cos\gamma) = - 1$.

 Together, equations (\ref{W} - \ref{g}) entail the following expression for the elevation in the submoon and antimoon points:
 \ba
 H\,=\,H_2\,+\,H_3\,+\,O(\s (R/r)^{5} \s)\,\;,
 \label{H}
 \ea
 where
 \ba
 H_2\,=\;h_2\;\frac{R^{\s 2}}{G\s M}\;\frac{G\,M^{\,\prime}}{r}\,\left( \frac{R}{r} \right)^2\,=\;h_2\,\frac{M^{\,\prime}}{M\,}\s\left( \frac{R}{r} \right)^3\s R\,\;,
 \label{elevation}
 \label{H_2}
 \ea
 \ba
 H_3=\,\pm\,h_3\,\frac{R^{\s 2}}{G\s M}\;\frac{G\,M^{\,\prime}}{r}\,\left( \frac{R}{r} \right)^3=\,\pm\,h_3\,\frac{M^{\,\prime}}{M\,}\s\left( \frac{R}{r} \right)^4\s R\,\;,
 \label{H_3}
 \ea
 the ``plus'' and ``minus'' signs corresponding to the sub- and antimoon points.

 For a moon on a circular synchronous orbit, $r$ is the synchronous radius:
 \ba
 r\,=\,
 \left(\frac{G\,(M + M^{\,\prime})}{ \Omega^{\,2} }\right)^{\s 1/3}\,\;,
 \label{r}
 \ea
 $\Omega$ being the planet's `initial' rotation rate. By `initial', we understand the rate established at the time of rotation synchronisation, i.e., equal to the mean motion of the moon at that time.
 This rotation rate was thereafter sustained by the synchronous moon till its elimination.

  Since a part of the centrifugal force mimics the quadrupole component of the tidal force (see Appendix \ref{centrifugal}),
  a link exists between the then values of the dynamical oblateness $J_2$, the Love number $k_2$, and the rotation rate $\Omega\;$:
 \ba
 \Omega^{\s 2}\s=\;\frac{3\, G\s M\s J_2}{R^{\,3}\, k_2}\;\,,
 \label{bills}
 \ea
 where $J_2$ is related to the principal moments of inertia $A<B<C$ through
 \ba
 J_2\s=\,\frac{C\s-\s(A\s+\s B)/ 2}{M\;R^{\,2}}\;\;.
 \label{J2}
 \ea

 Combined, formulae (\ref{r}) and (\ref{bills}) yield a convenient expression for the synchronous radius' early value:
 \ba
 r\,=\,R\,\left(\frac{k_2}{3\s J_2}\;\frac{M+M^{\,\prime}}{M}  \right)^{\,1/3}\;\,.
 \label{int}
 \ea
 A subsequent insertion of this expression in equations (\ref{H_2} - \ref{H_3}) results in
 \ba
 H_2\,=\;
 3\,J_2\,\frac{h_2}{k_2}\;   \frac{M}{\s M + M^{\,\prime}\s}\;\frac{\s M^{\,\prime}\s}{M\,}\; R\,=\,5\,J_2\;\frac{M^{\,\prime}}{\s M + M^{\,\prime}\s}\,R\;\;,
 \label{H2}
 \label{10}
 \ea
 \ba
 \nonumber
 H_3 &=&
 \pm\;h_3 \,\left(\frac{3\s J_2}{k_2}\;\frac{M}{M+M^{\,\prime}}  \right)^{\,4/3}\frac{\s M^{\,\prime}\s}{M\,}\;R
 ~\\
 \label{H3}\\
 &=&\pm\;H_2\;
 \frac{7}{3}\;\s
 \frac{k_3}{k_2^{4/3}}\;\s(3\s J_2)^{1/3}\s\left( \frac{M}{M+M^{\,\prime}} \right)^{1/3}
 \,\;,
 \nonumber
 \ea
 where we used the hydrodynamic relations $h_2/k_2=5/3$ and $h_3/k_3=7/3$ acceptable for a hot planet (Appendix \ref{A2}).

 We observe that the relative asymmetry $H_3/H_2$ is proportional to the factors $(3\s J_2)^{1/3}$ and $\s{k_3}\s{k_2^{\s-\s 4/3}}\s$.
 As will be explained in Section \ref{Numbers}, the former of these factors has undergone a very limited time evolution and may be attributed its present-day value: $\s(3\s J_2)^{1/3}=0.18\,$. By distinction, $\s{k_3}\s{k_2^{\s-\s 4/3}}$ dropped at some point. Thence, very naturally, the earlier the triaxial figure was formed the larger its seed asymmetry was. We shall dwell on this in Section \ref{Asymmetry} in more detail.

 \section{Parameters \label{Numbers}}

 Provided in Table 1, the present values of the Martian mass and mean radius,
 \ba
 M^{\rm(present)}\s=\,6.4169 \times 10^{23}~\mbox{kg}~~,\quad R^{\rm(present)}\,=\,3.3895 \times 10^{6}~\mbox{m}~~,
 \label{}
 \ea
 include the late veneer carried out mostly by the planetesimals left-over from the terrestrial planet formation period.
 According to \citet[Fig. 17a]{Morbidelli} it added 1.6\% to Mars's mass, almost a third of this addition having arrived  during the LHB.
 Hence the estimates for a young Mars:
 \ba
 M\s=\s 6.31 \times 10^{23}~\mbox{kg}~~,\quad R\s=\s 3.37 \times 10^{6}~\mbox{m}~~.
 \label{radius}
 \ea

    Now, the dynamical oblateness and triaxiality, whose current values are given in Table 1.
     With Tharsis ``removed'', $J_2$ is reduced by approximately $5 \%$ for a fully hydrostatic $J_2$ to approximately $10\%$ for $J_2$ that is $20\%$ nonhydrostatic \citep{zuber}. In our case, however, the reduction of $J_2$ would have been much less, because we would ignore not the entire Tharsis but only its highest part caused by later uplifts and eruptions, and would leave the tidal contribution. Another change in $J_2$ might have come from the accretion of exterior material, especially during the LHB.  While the arrival rate of deposits may have, in principle, depended on the latitude, any such nonuniformity of mass influx was smeared by the equinoctial precession of Mars and large variations of Mars' obliquity over its history. This averages out the influence of bombardment on $J_2$.
   These considerations legitimise our use of the present oblateness in calculations pertaining to the figure-formation stage
   (especially in formulae (\ref{H3}) and (\ref{relative}) for the seed triaxiality, where $J_2$ is raised to the power of $1/3$):
 \ba
 J_2\s\approx\,J_2^{\rm (present)}\s=\,1.96\times 10^{-3}\,\;.
 \label{oblateness}
 \ea

 Related to the principal moments $\s A<B<C\s$ through
 \ba
 J_{22}\s=\,\frac{B\s-\s A}{4\s M\s R^{\,2}}\;\;,
 \label{J22}
 \ea
 the dynamical triaxiality may have been more sensitive to later uplifts and volcanism, especially at Tharsis and opposite to it. (Recall that within our hypothesis the areas around the frozen tidal bulges are supposed to have been more prone to these processes.) So the relative increase of $J_{22}$ may have been more noticeable than that of $J_2$. Therefore the present triaxiality $J_{22}^{\rm (present)}$ serves as the upper bound for the early value $J_{22}\,$:
 \ba
 J_{22}\s \lesssim \;J_{22}^{\rm (present)}\,=\,6.31\times 10^{-5}\,\;.
 \label{tria}
 \ea

 \section{How massive a moon is needed?\label{mass}}

 For the purpose of estimate, we assume the seed equatorial asymmetry to be small, $\s|\s H_3\s|\s\ll\s H_2\s$,
 and model the soft young planet with a homogeneous ellipsoid having the principal axes $a>b>c$, whence
   $\s A\s =\s\frac{\textstyle M}{\textstyle 5}\s(b^2+c^2)
 \s$ and $\s B\s =\s\frac{\textstyle M}{\textstyle 5}\s(c^2+a^2)
 \s$. Insertion of these formulae in equation (\ref{J22}) produces
 \ba
 J_{22}\s=\s\frac{1}{5}\;\frac{a+b}{2\,R}\;\frac{a-b}{2\,R}\s\approx\s\frac{1}{5}\;\frac{a-b}{2\,R}\;\,.
 \label{}
 \ea
 Identifying the tidal elevation $H_2$ with $\frac{\textstyle a-b}{\textstyle 2}$, we obtain:
 \ba
 H_2\approx\, 5 \, R \, J_{22}\;\,,
 \label{14}
 \ea
 which is about $1.1$ km.

 Using formulae (\ref{10}) and (\ref{14}), we arrive at a remarkably simple relation:
 \begin{equation}
  \boxed{
 \begin{array}{rcl}
 M^{\,\prime}\,\approx\,\frac{\textstyle J_{22}}{\textstyle J_2}\, M\,\;.
  \end{array}
 }
 \label{ratio1}
 \end{equation}
 In this expression, both $J_2$ and $J_{22}$ are parameters of a young Mars. As was explained in Section \ref{Numbers}, the present value
 of the oblateness is still a tolerable estimate for the early oblateness, while the present triaxiality is only an upper bound for
 the early value. Combined with equality (\ref{oblateness}) and inequality (\ref{tria}), the above relation becomes:
 \begin{equation}
 \boxed{
 \begin{array}{rcl}
 M^{\,\prime}\,\lesssim\,\frac{\;\;\textstyle J_{22}^{\rm(present)}\;}{\;\textstyle J_2^{\rm(present)}\;}\; M
 \,\;.
 \end{array}
 }
 \label{ratio}
 \end{equation}
 %
 %
 Insertion of the numerical values of $J_{22}^{\rm (present)}$,  $J_2^{\rm (present)}$, \,and $M$ gives us:
 \ba
 M^{\,\prime}\,\lesssim\;3.2\times 10^{-2}\,M\,\approx\,2.0\times 10^{22}\;\mbox{kg}\;\,,
 \label{estimate}
 \ea
 which is less than one third of the lunar mass. It could have been even less than that, depending on how $J_{22}$ has been boosted by the later volcanic activity.

 Had we assumed that the elevation $H_2$ is comparable to the height of Syrtis Major Planum (on average, $2.1$ km) or of Tharsis (on average, about $7$ km above the reference datum), we would have raised the value of $H_2$ by a factor of 1.9 to 6.4, which would accordingly require, through equation (\ref{10}), a moon of mass $ (4.3~-~17)\times 10^{22}$ kg, which would be between $0.6$ and $2.3$ lunar masses.
 Geophysical evidence, however, is indicating that both Syrtis Major and Tharsis owe much of their elevation to
 processes subsequent to the initial tidal distortion (Syrtis assumed its present shape in the early Hesperian; Tharsis in the late Hesperian, if not later). So a third of the lunar mass, equation (\ref{estimate}), is a trustworthy bound for $M^{\,\prime}$.

 \section{A\s\,tidal\s\,origin\s\,of\s\,Mars'~equatorial~asymmetry\label{Asymmetry}}

 Owing to equations (\ref{H3}), (\ref{oblateness}), and (\ref{estimate}), the seed elevations  $H_2 + H_3$ in the submoon and antimoon points were
 \ba
 H_2\s\left[\s
 1\s\pm\,\frac{7}{3}\;
 \frac{k_3}{k_2^{4/3}}\;(3\s J_2)^{1/3}\s\left( \frac{M}{M+M^{\,\prime}} \right)^{1/3}
 \s\right]\s=\s H_2\s\left[\s
 1\s\pm\s 0.416\,
  \frac{k_3}{k_2^{4/3}}
 \s\right]\;\,,\;\;
 \label{relative}
 \ea
 with the ``plus'' sign for the submoon point, ``minus'' for antimoon.

 While addressing the timing issue would require modeling of the solidification of an early magma ocean, and of the formation and thickening of Mars' stagnant lid, a simplified analytical approach is nonetheless possible. The insertion of expression (\ref{kl}) for $k_l$ into the second term of equation (\ref{relative}) renders the relative triaxiality, $H_3/H_2$, as a function of $\mu$. This effectively would be a function of the figure formation time, had we known the history of solidification expressed as $\mu(t)$. Establishing of this time-dependence would constitute a separate, heavily numerical project. So for now we are using the evolving $\mu$ as a parameter, instead of time. Along these lines, we find in Appendix \ref{A2} that the values of  ${k_3}{k_2^{-4/3}}$ are residing in the interval
 \ba
 7.59\times 10^{-2}\,<\; {k_3}\s{k_2^{\s-\s 4/3}}  \;<\,3.32\times 10^{-1}\,\;.
 \label{ine}
 \ea
 The upper bound corresponds to a figure that was formed at the magma-ocean stage and sustained by the moon into the solidification epoch. This scenario implies that
 the moon showed up on the then synchronous orbit when the Martian mean shear modulus was not very different from that of a near-liquid lava, $\s\mu\s\simeq\s 0.2\times 10^9$ Pa.  The lower bound in equation (\ref{ine}) pertains to a scenario wherein the moon got synchronised when the solidification was already going on, with the mean viscosity about $\mu\simeq 18\times 10^9$ Pa.

 We observe that an early formation renders a higher seed asymmetry, with an elevation of $\simeq 1.14\,H_2$ in the submoon point, and $\simeq 0.86\, H_2$ in the antimoon one. A later formation produces a lower asymmetry, with elevations of $\simeq 1.03\,H_2$ and
  $\simeq 0.97\, H_2$ in the submoon and antimoon points, correspondingly. This may be regarded as an argument in favour of the moon having emerged and synchronised its mean motion during the magna-ocean period already. On the other hand, if part (b) of our hypothesis in Section \ref{Hypothesis} is correct, and the tidally elevated areas were most prone to geophysical processes, then even a small initial tidal asymmetry could later entail a larger asymmetry of shape~---~an issue requiring numerical treatment.

 \section{Sanity checks}

 \subsection{Nerio was staying outside the Roche sphere}


 For the synchronicity radius $r$ of a young Mars, equations (\ref{int}) and (\ref{oblateness}) yield:
 \ba
 \frac{r}{R}\;=\;5.56\;\sqrt[3]{k_2}\,\;,
 \label{inta}
 \ea
 wherefrom
 \ba
 %
 %
 3.88  \,<\,  \frac{r}{R}  \,<\, 6.22
 \;\,.
 \label{double}
 \ea
 Here the upper bound corresponds to a situation where the synchronous moon appeared and formed the figure already at the magma-ocean period,
 when the planet's mean rigidity $\mu$ was of order $0.2$ GPa. The lower bound corresponds to a slightly later formation when the solidification was starting, and the mean rigidity was about $18$ GPa.  Borrowed from a study of viscoelastic behaviour of basaltic lavas
 near the softening point \citep{James}, these limiting values define an interval for the of values of $k_2$, see Appendix \ref{A2}.
 The uncertainty in our knowledge of the interval for $\mu$, and therefore of the interval for $k_2$, is mitigated by $k_2$ being raised to the power of $1/3$ in the expression above. Still, the second decimal in the double inequality (\ref{double}) should be taken with caution, both because of the uncertainty of the limits on $\mu$ and because of the approximate nature of the assumption (\ref{oblateness}) for $J_2$.

 Even the minimal value of $r$ in equation (\ref{double}) is well above the Roche radius, which is
 \ba
 r_R =\, 2.20\,R\,\left(\frac{\rho_p}{\rho_m}\right)^{1/3}\,=\,2.20\,R\,\left(\frac{\rho_p}{\rho_M}\right)^{1/3}
   = 2.32 R
 \,\;.
 \label{roche}
 \ea

 In this expression for $r_R$, the average density of the planet Mars is $\rho_p = 3934 $  kg m$^{-3}$, while the average density $\rho_m$ of the moon is approximated with that of our own Moon, $\rho_M =3344$ kg m$^{-3}\s$. For the overall factor, we employed not Chandrasekhar's incompressible-fluid factor $2.46$ but a lower value $2.20$ appropriate to rubble piles with some shear strength \citep{Leinhardt}. The actual factor for solids should be even smaller, which would further ensure the desired inequality $\s r>r_R\s$.

 While Phobos, with its $a = 2.76\s R^{\rm (present)}$, is now entering the Roche sphere whose radius is $2.80\s R^{\rm (present)}$, for Nerio equation (\ref{roche}) gives $2.32$ instead of $2.80$ because the density of a large moon is much higher than that of Phobos.

 \subsection{Nerio's mass was sufficient\\ to synchronise Mars' rotation}


 \subsubsection{Scenario 1. Tidal synchronisation by a receding moon}

  If the moon is tidally receding, synchronisation may be achieved if several conditions are fulfilled. The synchronous radius
  $
     r
  $
  must be residing between the Roche radius $r_R$ and the reduced Hill radius $\s r_H^{\,\prime}\s\;$:
  \ba
  r_R\s<\;
     r
  \,<\;r_H^{\,\prime}\;\,.
  \label{1}
  \ea
  For dynamical consequences of these two inequalities, see \citet[Appendix A and Section 5, correspondingly]{pathways}. Neither of these inequalities imposes a restriction on the moon-to-planet mass ratio $M^{\,\prime}/M\s$.

  The third condition to be obeyed is this: the synchronism must be attained before the moon leaves the reduced Hill sphere. Most counterintuitively, this condition is not the same as
  $
     r
  \,<\;r_H^{\,\prime}\s$.  What it actually implies is that at some instant of time (not necessarily from the start) the expansion rate of the synchronous radius must exceed the expansion rate of the moon's orbit~---~and must keep exceeding it until the synchronisation.  This condition does render a restriction on $M^{\,\prime}/M$. By equation (52) from {\it Ibid}, it reads:
  \ba
  \frac{M^{\,\prime}}{M}\;>\;\left(\s\frac{R}{r_H^{\,\prime}}\s\right)^2\;\,,
  \label{2}
  \ea
  $R$ being the planet's radius, and $r_H^{\,\prime}$ the reduced Hill radius given by equation (2) from {\it Ibid}. The insertion of parameters' values shows that this constraint is mild:
  \ba
  \frac{M^{\,\prime}}{M}\;>\;4.1\times 10^{-5}\,\;,
  \label{3}
  \ea
  and is easily satisfied by the moon we are having in mind.

 \subsubsection{Scenario 2. Tidal synchronisation by a descending moon\label{descending}}

 All said above about the two inequalities (\ref{1}) remains in force for tidal descent. Also, equation (\ref{2}), again, renders a third condition. As explained in \citet[Section 9.2]{pathways}, violation of this inequality leaves the moon no possibility to evade spiralling onto the Roche limit. So we again end up with the mild constraint (\ref{3}).


 \subsection{Timescales}

 \subsubsection{Partners' spin evolution}

 At the moment of its accretion or capture, the moon's rotation was likely faster than orbiting; and the timescale $\tau_m$ of evolution of its rotation may be termed simply ``despinning timescale'' or ``deceleration timescale''. For the planet, however, two options are possible. If the partners were tidally receding from one another, like the Earth and the Moon, the planet's rotation was slowing down; so the corresponding timescale $\tau_p$ can be set positive and named ``despinning timescale'' or ``deceleration timescale''. By distinction, were the synchronism being attained via tidal approach, the planet's rotation was accelerating. While the corresponding timescale $\tau_p$ still may be termed ``despinning timescale'' or ``deceleration timescale'', its sign should be set negative.
 As explained in Appendix \ref{timescales1}, both timescales $\tau_m$ and $\tau_p$ were short as compared to Mars' geological evolution times.

 For an impact-created moon, even if we assume the initial separation to be as large as $6\s R$ and both $Q^{\,\prime}/k_2^{\,\prime}$ and $Q/k_2$ as large as $10^3$, the resulting rotation evolution timescales will be several years for the moon, and about a hundred thousand years for the planet. This is two orders of magnitude shorter than the duration of the magma-ocean era for Mars.

 For a captured-moon scenario, choosing again both $Q^{\,\prime}/k_2^{\,\prime}$ and $Q/k_2$ to be as high as $10^3$, we find that the rotation-evolution timescales are short by the geological measure, assuming the initial separation did not exceed $\simeq 22\s R\s$. Specifically, for $a_0= 10\s R\s$ and $a_0= 22\s R\s$ , the planet's timescale $\tau_p$ turns out to be about $10^6$ yr and $ 3\times 10^7$ yr, correspondingly. The former value is much shorter than the duration of the magma-ocean stage, the latter one is comparable to that duration.

 Our estimates are very conservative. The actual timescales were shorter owing to the partners' tidal heating leading to lower values of $Q^{\,\prime}/k_2^{\,\prime}$ and $Q/k_2$.
 Still, a relative proximity of $\tau_p$ to the duration of the magma-ocean stage may, arguably, be regarded as a limitation on the captured-moon scenario:
 the capture should not have happened at too remote a distance. Detailed numerical model of a combined orbital and geophysical history is needed to say more on this.

 \subsubsection{Circularisation}

 Estimates for the circularisation timescale are derived in Appendix \ref{circ}. Until a mutual synchronism is attained, the tides in the already synchronised moon are working to decrease the eccentricity, while the tides in the still nonsynchronous planet are working to boost the eccentricity value. Therefore, circularisation cannot even begin until full synchronism is reached. Once the synchronism is established, circularisation is taking place over a timescale $\tau_e$ of about $0.1$ to $0.3$ of the tidal deceleration timescale $\tau_p$ of the planet. Therefore, there is a period of time after the synchronisation, during which the orbit retains its eccentricity. Geologically, this period is short. Whether it is sufficient to produce in the planet an amount of tidal heat sufficient to influence geophysical processes requires a detailed numerical study. As we mentioned above, the pattern of this heat production was surely inhomogeneous, with its maxima around the tips of the figure.

 Also, as pointed out in Appendix \ref{circ}, the circularisation was never complete because of two physical factors: the gravitational pull of the Sun and the role of Mars' triaxiality (recall that circularisation begins after the synchronism is attained~---~i.e., after a seed triaxiality is acquired by Mars). The Sun's gravity is especially important if the moon is tidally receding and crossing the 2:1 MMR with the Sun, at $a=3.2\s R$. This crossing entails an eccentricity jump of $0.0085$.  This eccentricity value becomes stable if the synchronous orbit happens to coincide with this MMR. The triaxiality of the planet plays an especially big role if the
 moon crosses the 2:1 resonance with Mars' figure, at $a=3.8\s R\s$. This crossing gives the eccentricity a jump of $0.032\s$~---~which would become a stable value for $e$ were the synchronous orbit able to coincide with this resonance. Though in a slight violation of inequality (\ref{double}), these resonances should not be excluded from consideration, given the uncertainty in our knowledge of the then value of $k_2$ showing up in equation (\ref{inta}).

 \subsection{Synchronism as Mars' spin-orbit end-state\label{end-state}}


  As was pointed out above, circularisation becomes efficient only after synchronisation. In other words, prior to the synchronisation, the binary could have kept its eccentricity high, and could have been even boosting it, depending on the frequency-dependencies of the quality functions $k_2/Q$ and $k_2^{\prime}/Q^{\,\prime}$ of the planet and the moon. Therefore, during the spin evolution of the planet, various spin-orbit resonances were crossed by the system. As was explained in detail by \citet{Noyelles}, such a system can under some circumstances get stuck in one or another higher spin state~---~like Mercury did in the course of its tidal despinning.

  It raises an issue if Mars also could have been captured in a higher spin state in the course of the Mars-Nerio spin-orbit evolution. This is important, because once a partner is caught in a higher spin state, it may stay there for good, or until some event either separates the pair or somehow pushes it out of this spin state. The potential possibility of Mars' entrapment in a higher spin state may therefore portend a difficulty for our theory. The question then becomes if such a possibility is realistic. An answer to this question hinges on several parameters, including the eccentricity value. E.g. Mercury would have not been trapped into a higher spin state had its high eccentricity not been sustained by other planets' pull \citep{Noyelles}.

  Born {\it in situ}, or created by impact, or captured in the disk, a moon had a low initial eccentricity, if any. So both partners despan directly to synchronism.  Also, a hot semimolten rotator cannot get stuck in a higher spin-orbit state. Moreover, had it somehow been captured in such a state during the short period between the accretion and the formation of a magma ocean, it would definitely have slipped out of that spin state during the magma-ocean stage, as explained in detail in Appendix~\ref{synchronism}.

 \section{Results and questions}

  It is proposed in this work that an initial, `seed' triaxiality of Mars was produced by a massive moon that was orbiting a young Mars on a synchronous orbit. Depending on whether the moon's orbit was circular or not, the tide was static or near-static (slightly librating about the mean direction toward the synchronous moon). After the seed triaxiality was created, the tidally elevated areas on the two opposite tips became, hypothetically, more prone to geophysical processes than the rest of the Martian surface~---~which led to these tips' further rise and a resulting triaxiality increase.

 \subsection{Results}

 \enlargethispage{\baselineskip}

  The principal result of our study is that a moon of less than a third of the lunar mass was capable of generating the seed triaxiality. This result is deduced from the following two requirements:
  \begin{itemize}
  \item[(1)~] The moon must have synchronised its mean motion with Mars' rotation when Mars was still sufficiently hot, its lithosphere being absent or weak, and its tidal response being predominantly hydrostatic.
  \item[(2)~] The present-day triaxiality $J_{22}^{\rm (present)}$ is the upper boundary for the value $J_{22}$ tidally caused by the putative moon.
 \end{itemize}
  The putative moon Nerio failed to survive the LHB.  Phobos and Deimos may be remnants of its large fragment, which was destroyed later. Indeed, while the LHB occurred 4.1 to 3.8 Gyr ago, a probable common ancestor of Phobos and Deimos was destroyed by an impact between 1.6 and 2.7 Gyr ago \citep{phobosdeimos}~---~which eliminates the possibility of Phobos and Deimos being Nerio's {\it immediate} progeny.

  A symmetrical part of the seed triaxiality having originated due to the quadrupole component of the tide, the seed asymmetry was produced by the octupole component. The relative magnitude of this initial asymmetry depended on the time of creation of the initial figure.
  If the synchronous moon had already existed during the magma-ocean stage of Mars, the seed asymmetry would comprise an additional 14\% elevation in the submoon point and a 14\% reduction of elevation in the antimoon point, as compared to the symmetrical part of the elevation. If, however, the moon was synchronised somewhat later, when the process of solidification began but a lithosphere was not yet fully formed, then the seed asymmetry would be at a several percent level only.  It remains to be studied by numerical modeling if the current asymmetry between Tharsis and its opposite province is due to the initial tidal distortion~---~or if the initial disparity was later boosted by geological processes in the submoon and antimoon zones (like the emergence of plumes).

 \subsection{Questions}

  Our paper serves as a motivation to numerically explore the evolution of a young planet whose shape is neither spherical nor even elliptic.  At large, the question is if a limited initial asymmetry of the shape can entail a considerable asymmetry in the process of solidification and in the subsequent geological activity~---~convection, uplifts, volcanism.

  More specifically, is it right to assume that geological processes were more intense in the tidally elevated areas?  This question emerges both for a moon on a circular synchronous orbit (with no tidal heating in the planet) and for a moon describing an eccentric synchronous orbit and generating additional heat in the young Mars.  A study by \citet{segatz88} has demonstrated that in a body (Io) synchronised by its partner (Jupiter) tidal heating is inhomogeneous.  Whether the maximally heated areas are in the sub- and anti-Jovian zones is not warranted and depends on the internal structure. Also, that study was carried out for a spherical body and in the quadrupole approximation only.  In our case, the nonspherical (and nonellipsoidal) shape of the tidally distorted young Mars will be crucial. Also, due to the proximity of the perturber, the degree-3 inputs into the tidal dissipation rate will have to be included. A study of the kind, carried out by \citet{Quillen} for the Moon, offers to our theory both a {\it pro} and a {\it contra} argument. On the one hand, the degree-3 component of the tide generates a considerable difference in the tidal heat flux between the near and far sides, the near side being predictably warmer and therefore having better chances for subsequent uplift and volcanism. On the other hand, within the considered model the heat flux increases toward the polar zones, thus posing a question on why the volcanic activity near the poles has been weaker than on Tharsis. The model by \citet{Quillen} relying on the Kelvin-Voigt rheology, a more accurate study based on a realistic rheology is then required.

  Had the moon disintegrated in the early LHB, there should have been an enhanced impactor flux at low latitudes on Mars. Would the subsequent LHB, along with geological processes, be sufficient to wipe out the resulting latitudinal dependence of the most ancient crater population?
  As of now, the oldest crater populations show no such dependence.

   \enlargethispage{\baselineskip}

 \section*{Acknowledgments}

 The author wishes to acknowledge helpful consultations provided to him by
                      Amirhossein Bagheri,
   Andrew Dombard,
   Amir Khan,
                      Alessandro Morbidelli, Anthony L. Piro,
              Robert Schulmann,
                      Francis Nimmo
                      and
   Michaela Walterova,
              none of which colleagues shares responsibility for the hypothesis proposed in this work.

 The author was also fortunate in having had the assistance of two reviewers who donated generously of their time, helping to shape and improve the contours of the manuscript. Their advice is greatly appreciated. Particular thanks are owed to Beno\^{i}t Noyelles for suggesting the name Nerio for the putative moon.

     \vspace{1mm}
   \noindent   This research has made use of NASA's Astrophysics Data System.\\

 \section*{Open Research}

 The shear rigidity values used in this paper are borrowed from \citet{James}. The value of Mars' present dynamical oblateness $J_2^{(present)}$ is provided in \citet{Konopliv_2011}.  The value of Mars' present dynamical triaxiality $J_{22}^{(present)}$ is given in \citet{Konopliv}.
 The value $M^{(present)}$ of Mars' present mass is borrowed from the
 \href{https://pds-geosciences.wustl.edu/mro/mro-m-rss-5-sdp-v1/mrors_1xxx/data/shadr/jgmro_120d_sha.lbl}{PDS file for the Mars gravity field MRO120D} used in \citet{Konopliv2016}.

~\\

   \appendix

   \noindent
   {\Large{APPENDIX}}

 %
 %
 %

   \enlargethispage{\baselineskip}

 \section{Love numbers \label{A2}}

 During the magma-ocean stage and for some time thereafter, while the lithosphere was still weak, the expressions intended for a homogeneous body were applicable:
 \ba
 k_l\s=\,\frac{3}{2\,(l\s - \s1)}\;\frac{1}{1\s+\s{\cal B}_l\,\mu}\,\;,
 \label{kl}
 \ea
 \ba
 h_l\s=\,\frac{2\,l\s +\s 1}{2\,(l\s -\s 1)}\;\frac{1}{1\s+\s{\cal B}_l\,\mu}\,\;,
 \label{hl}
 \ea
 \ba
 {\cal B}_l\,=\,\frac{3\,(2l^2+\s 4\, l\,+\,3)}{4\,l\,\pi\, G\,(\rho\, R)^2}\;\;.
 \label{Bl}
 \ea
 Derived, historically, for homogeneous elastic spheres, these formulae can be used also for homogeneous viscoelastic spheres under static loading,
 as can be observed from expressions (19) and (44) in \citet{Efroimsky2015}. 
 
 From equations (\ref{kl} - \ref{hl}), ensue two simple formulae used in Section \ref{Formulae}:
 \ba
 \frac{h_2}{k_2}\;=\;\frac{5}{3}\;\;,\qquad \frac{h_3}{k_3}\;=\;\frac{7}{3}\,\;.
 \label{}
 \ea
 For a young Mars, the dimensional quantities ${\cal B}_2$ and ${\cal B}_3$ assume the values
 \ba
 \quad {\cal B}_2\,=\,0.193\times 10^{-9}\;\mbox{kg}^{-1}\,\mbox{m}~\mbox{s}^{2}\;\,,\,\quad {\cal B}_3\,=\,2.24\times 10^{-9}\;\mbox{kg}^{-1}\,\mbox{m}~\mbox{s}^{2}
 \,\;.
 \label{B}
 \ea
 Studies of viscoelastic behaviour of basaltic lavas demonstrate that on approach to a sample's softening point its shear rigidity $\mu$ is of order $\simeq 18\times 10^9$ Pa, but goes down to $\simeq 0.2\times 10^9$ Pa as sample fluidity begins to rapidly increase at temperatures exceeding $1100^{\,\rm o}$ C \citep{James}:
 \ba
 \mu\,=\,(0.2 - 18)\times 10^{9}\;\s\mbox{Pa}\,\;.
 \label{mu}
 \ea
  This indicates that the Love numbers at the earliest stages were residing within intervals whose upper bounds were close to the ideal-fluid values:
 \ba
 k_2\,=\,0.34\;-\;1.4\;\;,\;\quad h_2\,=\,0.56\;-\;2.4\,\;,
 \label{k2}
 \ea
 \ba
 k_3\,=\,0.018\;-\;0.52\;\;,\quad h_3\,=\,0.042\;-\;1.2\,\;.
 \label{}
 \ea
 The value of each of these Love numbers was at the upper bound of the corresponding interval during the magma-ocean stage (with $\mu=0.2\times 10^{9}$ Pa),
 and was descending to the lower bound during the solidification stage, as the rigidity was approaching $\mu=18\times 10^{9}$ Pa.

 For comparison, the present-day value of $k_2$ of the solid Mars is residing between $0.169$ and $0.174$ \citep{Konopliv}.

 In section \ref{Asymmetry}, we needed to evaluate the product $\s{k_3}\s{k_2^{\s -\s 4/3}}$. Using formulae (\ref{kl}) and (\ref{B}), we find
 \ba
 7.59\times 10^{-2}\,<\; {k_3}\s{k_2^{\s-\s 4/3}}  \;<\,3.32\times 10^{-1}\,\;,
 \label{}
 \ea
 the maximum corresponding an early figure formation ($\mu\simeq 0.2\times 10^9$ Pa), the minimum to slightly later formation, with solidification already going on
 ($\mu\simeq 18\times 10^9$ Pa).

 \section{Realistic values for the $k_2/Q$ of a young Mars\label{deta}}
 \setcounter{figure}{0}
 \renewcommand\thefigure{B\arabic{figure}}

 Figure S\s3\s{b} in the Supplement to Samuel et al (2019) suggests that, when Mars was 0.9 Byr old, its $k_2/Q$ was already as low as (0.22 - 0.24) $\times 10^{-3}\s$.
 We however should be cautious about this result, because it is based on the hypothesis of an early origin of the Martian moons, a hypothesis applicable to Deimos but not necessarily to Phobos. Integration of Phobos' orbit backwards in time demonstrates a rapid increase in its eccentricity \citep{phobosdeimos}
 This means that in the past the tides in Phobos were stronger than those in Mars, and dominated Phobos' descent. This indicates that the values of $k_2$ and $k_2/Q$ provided by Samuel et al (2019) should be used as the low bounds, at best.

\begin{figure}[H]
  \includegraphics[width=0.91\textwidth]{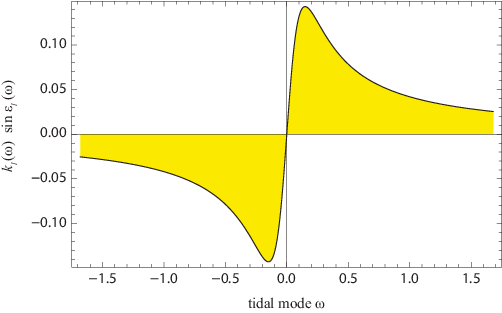}
             \caption{. \small A typical shape of the quality function $\,k_l(\omega)\,\sin\epsilon_l(\omega)\,$, ~where $\,\omega\,$ is a shortened notation for the tidal Fourier mode $\,\omega_{\textstyle{_{lmpq}}}\,$. \,(From Noyelles et al. 2014.)}
         \label{figure1}
     \end{figure}

 Also, the values of $k_2/Q$ from the said figure in Samuel et al (2019) pertain to a Mars of 0.9 Byr or older. We however are interested in a younger Mars, whose crust was only beginning to mature, but whose interior was still semimolten and not completely differentiated. Modeling this Mars with a homogeneous Maxwell body, we enquire about the magnitudes of and spread between the peaks in the tidal-mode-dependence of $k_2/Q\;$, see Figure \ref{figure1}.

\noindent
 As shown, e.g. in \citet[Section 4.4]{walterova}, the peak is residing at
 \ba
{\omega_{\rm{peak}}}\,=\;
 \;\frac{\tau_{\rm{_M}}^{-1}}{1\,+\,{\cal{B}}_2\,\mu}\,=\,
 \frac{\mu/\eta}{1\,+\,{\cal{B}}_2\,\mu}
 \label{wh}
 \ea
 and has a viscosity-independent magnitude
 \ba
 \left(\frac{k_2}{Q}  \right)^{\rm{(peak)}}\s=\;\frac{3}{4}\;\frac{ {\cal B}_2\,\mu }{\s 1 + {\cal B}_2\,\mu\, }\,\;.
 \label{see}
 \ea
 [\,Mind a misprint in \citet[Eqn 62]{Bagheri}.\s]

 The magnitude remains viscosity-independent also when the Burgers and Andrade terms show up as the mantle cools down, because these terms usually become prominent well to the right of the peak \citep{walterova}.

 For high Maxwell times (say, months or years, or longer), the peaks are sharp and stay close to zero. So the values of $k_2/Q$ observed
 at realistic forcing frequencies are much lower than the peak value, and are defined by how quickly the slope is falling off with the frequency increase.
 In fact, the slope of the tail does not fall off as quickly as shown in our figure, because at frequencies well above $\tau_M^{-1}$ transient processes kick in, and the rheology deviates from Maxwell towards Andrade (to be more exact, towards Sundberg-Cooper), see Figure 3 in \citet{walterova}.
 Most importantly, for a semimolten planet with a Maxwell time of order hours to days, this kink becomes spread widely; and at comparable periods the values of $k_2/Q$ are not dramatically different from the peak value. So, while the peak value (\ref{see}) may be by an order or two of magnitude higher than an actual $k_2/Q$ of a cold planet, it is a tolerable estimate for the $k_2/Q$ of a hot planet.

 Now, numbers.  In the extreme case of the magma-ocean stage, we have, from Appendix \ref{A2}: $\mu=0.2\times 10^9$ Pa, ${\cal B}_2\mu \approx 3.9\times 10^{-2}$ and therefore $\,\left({k_2}/{Q}  \right)^{\rm{(peak)}}\s\approx\s 0.28\times 10^{-1}\,$. If however we agree that the moon was giving
 Mars its shape shortly thereafter, i.e. when solidification began, then the corresponding value $\mu = 18\times 10^9$ Pa from Appendix \ref{A2} gives us ${\cal B}_2\mu \approx 3.5$ and therefore $\,\left({k_2}/{Q}  \right)^{\rm{(peak)}}\s\approx\s 0.58\s$. Thus we see that, for a young Mars, $k_2/Q$ is taking values, roughly, between
 $0.03$ and $0.6$.

 \section{Splitting the centrifugal force into a radial and quadrupole components \label{centrifugal}}

 Using the expression $\s P_2(\cos\gamma)=\frac{\textstyle 1}{\textstyle 2}\left(3\s\cos^2\gamma- 1\right)$, we split the centrifugal acceleration into a quadrupole and radial parts,
 \ba
 \nonumber
 \Omegabold\times\left(\Omegabold\times\eRbold\right)&=&\nabla\left[\frac{1}{2}\Omegabold^{\,2}\eRbold^{\,2}\left(\cos^2\gamma-1\right)\right]
 ~\\
 \label{}\\
 \nonumber
 &=&\nabla\left[\frac{1}{3}\Omegabold^{\,2}\eRbold^{\,2}
 \left[ P_2\left(\cos\gamma\right)- 1\right]\right]\;\,,
 \nonumber
 \ea
 the former influencing the geometric shape, the latter not. The quadrupole potential
 is compensated by the incremental tidal potential:
 \ba
 \frac{\textstyle 1}{\textstyle 3}\s\Omegabold^{\,2}\s\eRbold^{\,2}\s P_2\left(\cos\gamma\right)\s-\, k_2\s\frac{\textstyle G M}{\textstyle R}\,J_2\,P_2\left(\cos\gamma\right)\s=\s 0\,\;,
 \nonumber
 \ea
 whence expression (\ref{bills}).

 \section{Rotation rate evolution timescales\label{timescales1}}

 \subsection{Generalities}

  The tidal deceleration timescales of the moon and planet are, correspondingly,
 \ba
 \nonumber
 \tau_m\s=\;
 \frac{Q^{\,\prime}}{k_2^{\s\prime}}\;\frac{M^{\,\prime}}{M}\;\frac{M\s+\s M^{\,\prime}}{M}\;\frac{1}{n}\s\left(\frac{a}{R^{\,\prime}}  \right)^3
 & \approx &
 \frac{Q^{\,\prime}}{k_2^{\s\prime}}\;\frac{M^{\,\prime}}{M}\;\frac{1}{n}\s\left(\frac{a}{R^{\,\prime}}  \right)^3\\
  \nonumber\\
  \nonumber\\
 & = &
 \frac{Q^{\,\prime}}{k_2^{\s\prime}}\;\frac{M^{\,\prime}}{M}\,\frac{\rho_m}{\rho_p}\;\Pi  \qquad
 \label{m}
 \ea
 and
 \ba
 \tau_p\s=\;\pm\;
 \frac{Q}{k_2}\;\frac{M}{M^{\,\prime}}\;\frac{M\s+\s M^{\,\prime}}{M^{\,\prime}}\;\frac{1}{n}\s\left(\frac{a}{R}  \right)^3
 \s\approx\;\pm\;
 \frac{Q}{k_2}\;\left(\frac{M}{M^{\,\prime}}\right)^2\Pi  \;\,,
 \label{p}
 \ea
 $k_2/Q$ and $k_2^{\s\prime}/Q^{\,\prime}$ being the tidal parameters of the planet and moon, $\rho_p$ and $\rho_m$ being their densities,
 and a common dimensional factor $\Pi$ given by equation (\ref{product}) below.

 The ``$\pm$'' sign in the expression for $\tau_p$ implies that a positive despinning timescale $\tau_p$ corresponds to a setting where the synchronism is attained by the
 recession of the moon (like in the Earth-Moon case, with the Earth's rotation gradually decelerating). A negative despinning timescale is characteristic of an opposite situation where the synchronism is achieved by tidal descent of the moon and acceleration of the planet's rotation. On both occasions, one may use the terms ``despinning time'' or ``deceleration time'', keeping in mind that this ``despinning'' can sometimes be actually a spin-up.

 Expressions (\ref{m}) and (\ref{p}) follow, e.g. from equation (116) in \citet{Efroimsky2012}. An independent check can be performed by combining formulae (123), (132), and (188) from \citet{BoueEfroimsky2019}.
 Since \citet{soter} were interested in the despinning of the smaller partner, their equation (17) agrees with the r.h.s. of our expression (\ref{m}), not of (\ref{p}).

 Dividing formulae (\ref{m}) and (\ref{p}) by one another, we obseve that $\tau_m/\tau_p \ll 1$
 in all realistic situations. This is so owing to the smallness of the factor
 $\left(M^{\,\prime}/M\right)^3\left(\rho_m/\rho_p\right) < 3\times 10^{-5}\s$ emerging in that ratio.

 The common factor $\Pi$ entering formulae (\ref{m}) and (\ref{p}) is given by
 \ba
 \Pi = n^{-1}\left(\frac{a}{R}\right)^3\,=\;\frac{R^{\s 3/2}}{  \sqrt{G\s(M+M^{\,\prime})}}\,\left(\frac{a}{R}\right)^{9/2}
 \label{product}
 \ea
 and scales as $\s a^{4.5}$. E.g. for Phobos,
 it has the value $\Pi_{\rm \s Phobos} = 0.88\times10^5$~s.

 For the hypothetical moon Nerio, the lower bound on $\Pi$ pertains to a situation where $a_0$ (the initial value of $a$, one corresponding to a freshly formed or captured moon) equals the Roche radius $r_R=2.32\s R\s$. This yields:
 \ba
 \operatorname{min}\,\Pi\,=\,0.41\times 10^5\,\;\mbox{s}\,\;.
 \label{minpi}
 \ea
 In this case, the subsequent synchronisation of the orbit with the planet's rotation could have been achieved only through tidal recession, because tidal descent would have driven the moon into the Roche zone.  Keeping in mind that $\frac{\textstyle M^{\,\prime}}{\textstyle M}\,\frac{\textstyle \rho_m}{\textstyle \rho_p}\s\approx 2.7\times 10^{-2}\s$, we deduce that within this scenario the synchronisation of both partners is almost instantaneous. E.g. for $Q^{\,\prime}/k_2^{\prime}$ and $Q/k_2$ of order $10^3$,  the timescales come out as low as
 \ba
 \operatorname{min}\,\tau_m\,\simeq\,10^6\,\;\mbox{s}  \quad,\qquad \operatorname{min}\,\tau_p\,\simeq\,4\times 10^{10} \,\;\mbox{s}\,\;.
 \label{}
 \ea

 To find the maximal values of $\tau_m$ and $\tau_p$, we need to evaluate the upper bound on $\Pi$, one corresponding to the maximal realistic initial value $a_0$ of the semimajor axis. Two cases have to be addressed separately.

 \subsection{A moon formed by impact}

 In numerical simulations, most clumps left orbiting an Earth-sized planet after impact usually stay within several radii from its centre \citep{Ruiz}. To be on the safe side, assume that the initial semimajor axis obeyed $a/R < 6\s$; thence
 \ba
 \operatorname{max}\,\Pi\,=\,0.30\times 10^7\;\mbox{s}\;.
 \label{}
 \ea
 For $Q^{\,\prime}/k_2^{\prime}$ and $Q/k_2$ of order $10^3$, this entails the maximal values of
 \ba
 \operatorname{max}\,\tau_m\s\simeq\s 10^8\;\mbox{s}\;\simeq\, 3\s\;\mbox{yr}  \;\;\;,\qquad \operatorname{max}\,\tau_p\s\simeq\,3\times 10^{12}\;\mbox{s}\;\simeq\, 10^5\s\;\mbox{yr}\,\;,
 \label{sho}
 \ea
 i.e., the spin of the moon becomes synchronous within several years; of the planet within a hundred thousand years. Even if we (illegitimately) boost the values of $Q^{\,\prime}/k_2^{\prime}$ and $Q/k_2$ by another order of magnitude, we still shall be left with geophysically short times. In reality, the values of  $Q^{\,\prime}/k_2^{\prime}$ and $Q/k_2$
 should sooner be reduced than boosted, because both bodies were experiencing tidal heating. So the actual timescales of deceleration would be shorter than those given by equation (\ref{sho}).

  \subsection{A captured moon}

 In the disk, the moon could have been captured, in principle, at an initial separation exceeding that typical for creation of a moon by impact.
 The factor $\Pi$ and, consequently, the spin evolution times $\tau_m$ and $\tau_p$ scale as $a^{4.5}$. An increase of the initial separation $a_0$
 by a factor of $5/3\s\approx\s 1.67$ would result in a ten-fold increase of the time scales. For example, had we taken in the above calculation the initial separation to be $\s a_0=10\s R\s$, we would have obtained, instead of values (\ref{sho}), the values
 \ba
 \operatorname{max}\,\tau_m\s\simeq\s 10^9\;\mbox{s}\;\simeq\, 30\s\;\mbox{yr}  \;\;\;,\quad \operatorname{max}\,\tau_p\s\simeq\,3\times 10^{13}\;\mbox{s}\;\simeq\, 10^6\s\;\mbox{yr}\;,
 \label{sho2}
 \ea
 while for $\s a_0=22\s R\s$ it would have been
 \ba
 \operatorname{max}\,\tau_m\s\simeq\s 3.5\times 10^{10}\;\mbox{s}\;\simeq\, 10^3\s\;\mbox{yr}  \;\;\;,\quad \operatorname{max}\,\tau_p\s\simeq\, 10^{15}\;\mbox{s}\;\simeq\, 3\times 10^7\s\;\mbox{yr}\,.
 \label{sho3}
 \ea
 Larger initial values $a_0$ may not fit into our story because of wrong timing. Our scenario requires the planet to synchronise its rotation with the mean motion $n$ while the planet is still sufficiently deformable and capable of acquiring an asymmetric tidal shape to freeze later. We thus expect the synchronisation to happen well before the planet's age is $\sim 10^8$ years. Modeling by \citet[Fig S3b]{samuel} suggests that by the age of $\sim (2 - 8)\times 10^8$ yr the planet's $k_2/Q$ becomes comparable to its present value, indicating that the planet's shape may have started freezing by that time.

 Just like in the case of an impact-made moon, so in the case of a captured moon, tidal heating in the partners was working to increase the values of $k_2/Q$ and $k_2^{\prime}/Q^{\,\prime}$ and, consequently, to shorten the actual despinning times.

  \subsection{Conclusions}

  No matter how the moon was acquired, for an initial separation not exceeding $22\s R$ and the values of $Q^{\,\prime}/k_2^{\prime}$ and $Q/k_2$ not exceeding $10^3$, the spin evolution timescales are short by the geological measure. For an initial separation less than $10\s R$, synchronisation is especially swift.

  In reality, both timescales should have been shorter than those obtained above, because tidal heating reduces the values of both $Q^{\,\prime}/k_2^{\prime}$ and $Q/k_2$.

 \section{Tidal circularisation\label{circ}}

 When the spin of neither body is synchronised, while both obliquities (inclinations of the equators on the orbit plane) are small, the leading terms in the eccentricity rate are those linear in $\,e\,$ \citep{BoueEfroimsky2019}:
 \ba
 \nonumber
 \left(\frac{de}{dt}\right)_{l=2}
 =
  \,-\,n\,e\,\frac{\,M^{\,\prime}}{M\,}\,\left(\frac{R}{a}\right)^{\textstyle{^{5}}}\,
 \left[
 -\,\frac{3}{16}\,K_2(n-2\dot{\theta})\,-\,\frac{3}{4}\,K_2(2n-2\dot{\theta})
 \right.\\
 \nonumber\\
 \nonumber\\
 \nonumber
 \left.
 \,+
 \,
 \frac{147}{16}\,\,K_2(3n-2\dot{\theta})
 \,+\,\frac{9}{8}\,\,K_2(n)
 \right]\;\;\;\;\\
 \nonumber\\
 \nonumber
 -\,n\,e\,\frac{M\,}{\,M^{\,\prime}}\,\left(\frac{R^{\;\prime}}{a}\right)^{\textstyle{^{5}}}\,
 \left[
 -\,\frac{3}{16}\,K_2^{\,\prime}(n-2\dot{\theta}^{\,\prime})\,-\,\frac{3}{4}\,K_2^{\,\prime}(2n-2\dot{\theta}^{\,\prime})
  \right.\\
 \nonumber\\
 \nonumber\\
 \nonumber
 \left.
 \,+
  \,\frac{147}{16}\,K_2^{\,\prime}(3n-2\dot{\theta}^{\,\prime})
  \,+\,\frac{9}{8}\,K_2^{\,\prime}(n)
 \right]\\
 \nonumber\\
 \nonumber\\
 \;+\;O(e^2)\;+\;O(i^2)\,+\,O({i^{\,\prime\;}}^2)~~_{\textstyle{_{\textstyle .}}}\quad
 \label{53}
 \ea
 Here, $\dot{\theta}$ and $\dot{\theta}^{\,\prime}$ are the planet's and moon's rotation rates, while $K_2$ and $K_2^{\,\prime}$ are their (odd) tidal quality functions, such that $|K_2| = k_2/Q$ and $|K_2^{\,\prime}| = k_2^{\prime}/Q^{\,\prime}$.

 When both partners satisfy the Constant Phase Lag (CPL) model (so both $\,K_2\,$ and $\,K_2^{\,\prime}\,$ are constants) and  both $\,\dot{\theta}\,$ and $\,\dot{\theta}^{\,\prime}\,$ exceed $\,3n/2\,$, we have:
 $$K_2(n)\,=\,-\,K_2(n-2\dot{\theta})\,=\,-\,K_2(2n-2\dot{\theta})\,=\,-\,K_2(3n-2\dot{\theta})\,=\,k_2/Q\,$$
 and
 $$K^{\,\prime}_2(n)\,=\,-\,K_2^{\,\prime}(n-2\dot{\theta}^{\,\prime})\,=\,-\,K_2^{\,\prime}(2n-2\dot{\theta}^{\,\prime})\,=\,-\,K_2^{\,\prime}(3n-2\dot{\theta}^{\,\prime})\,=\,k_2^{\,\prime}/Q^{\,\prime}\,\;,$$
 wherefrom
 \ba
 \nonumber
 ^{\textstyle{^{(CPL)}}}\left(\frac{de}{dt}\right)_{l=2}
 =\,\frac{57}{8}\,n\,e\,\left[
 \frac{\,M^{\,\prime}}{M\,}\,\left(\frac{R}{a}\right)^{\textstyle{^{5}}}\,\frac{k_2}{Q}
 \;+\;
 \frac{M\,}{\,M^{\,\prime}}\,\left(\frac{R^{\;\prime}}{a}\right)^{\textstyle{^{5}}}\,\frac{k_2^{\,\prime}}{Q^{\,\prime}}
 \right]
   ~\\
   \label{CPL}\\
 \;+\;O(e^2)\;+\;O(i^2)\,+\,O({i^{\,\prime\;}}^2)~_{\textstyle{_{\textstyle .}}}\quad
 \nonumber
 \ea
 This is in agreement with \citet[Eqn (A1)]{Lainey} but differs from the corresponding formulae in some other works.

 When both partners satisfy the Constant Time Lag (CTL) model (i.e., when both $\,K_2\,$ and $\,K_2^{\,\prime}\,$ are linear in the tidal mode), expression (\ref{53}) becomes
 \ba
 \nonumber
 ^{\textstyle{^{(CTL)}}}\left(\frac{de}{dt}\right)_{l=2}
 =\;\frac{3}{4}\,n\,e\,\left[
 \frac{\,M^{\,\prime}}{M\,}\,\left(\frac{R}{a}\right)^{\textstyle{^{5}}}\,\frac{11\;\dot{\theta}\;-\;18\;n}{\dot{\theta}\;-\;n}
 \;K_2(2n-2\dot{\theta})
 \right.
 ~\\
\nonumber\\
 \left.
 \;+\;
 \frac{M\,}{\,M^{\,\prime}}\,\left(\frac{R^{\;\prime}}{a}\right)^{\textstyle{^{5}}}\,\frac{11\;\dot{\theta}^{\,\prime}\;-\;18\;n}{\dot{\theta}^{\,\prime}\;-\;n}
 \;K_2^{\,\prime}(2n-2\dot{\theta}^{\,\prime})\right]
   \label{CTL}\\
   \nonumber\\
 \;+\;O(e^2)\;+\;O(i^2)\,+\,O({i^{\,\prime\;}}^2)~~_{\textstyle{_{\textstyle .}}}\quad
 \nonumber
 \ea
 This agrees with \citet[Eqn 10]{Hut} and \citet[Eqn 19]{Emelyanov}, but not with \citet[Eqn 46]{Kaula} who neglected a factor of $\,(M+M^{\,\prime})/M\,$ and lost the factor of 4 in the denominator in the second line of his expression.

 We observe that within the CTL model the sign of a partner's input coincides with $\operatorname{Sign}\,(18n-11\dot{\theta})$. A body is mitigating the eccentricity $e$ when
 spinning swiftly ($\s\dot{\theta}/n > 18/11$), and is boosting $e$ when spinning slowly ($\s\dot{\theta}/n < 18/11$).
 This is in contrast to the CPL case, equation (\ref{CPL}), where each partner, whatever its spin, is always working to increase $e$.

 The sensitivity of $de/dt$ to the frequency-dependencies of $K_2$ and $K_2^{\,\prime}$ indicates that, even before one of the partners synchronises its spin,
 the eccentricity rate may have, in principle, been negative. However, for hot semimolten bodies (as well as for near-rubble ones) the rheology is not very different from CTL, because for such bodies interpeak interval in Figure \ref{figure1} is broad and includes the modes involved. So the tides in such a body, on approach to synchronism ($\s\dot{\theta}/n < 18/11$), are working to boost the eccentricity.

 As we saw in Appendix \ref{timescales1}, both bodies' rotation synchronises quickly, whereafter the tides in both bodies start working to reduce the eccentricity.

After both partners are synchronised ($\dot{\theta} = \dot{\theta}^{\,\prime}=n$), the terms with
$\s K_2(2n-2\dot{\theta})\s$ and $\s K^{\,\prime}_2(2n-2\dot{\theta}^{\,\prime})\s$ in formula (\ref{53}) vanish, and we end up with


 \bs
 \ba
 \nonumber
 \left(\frac{de}{dt}\right)_{l=2}
  =
 & - & \frac{21}{2}\,n\,e\,\frac{\,M^{\,\prime}}{M}\,\left(\frac{R}{a}\right)^{\textstyle{^{5}}}\,
 K_2(n)
 ~\\
 \label{56}\\
 \nonumber
 & - & \frac{21}{2}\,n\,e\,\frac{M}{\,M^{\,\prime}}\,\left(\frac{R^{\,\prime}}{a}\right)^{\textstyle{^{5}}}\,
 K_2^{\,\prime}(n)
 \;+\;O(e^2)\;+\,O(i^2)\;+\,O({i^{\,\prime\;}}^2)~~_{\textstyle{_{\textstyle ,}}}\qquad
 \ea
 where we took into consideration that both $\s K_2\s$ and $\s K^{\,\prime}_2\s$ are odd functions.

 Here $\s K_2(n) = k_2(n)/Q(n)\s$ and $\s K^{\,\prime}_2=k_2^{\prime}(n)/Q^{\,\prime}(n)\s$. Therefore,
 \ba
 \nonumber
 \left(\frac{de}{dt}\right)_{l=2}  =
 & - & \frac{21}{2}\,n\,e\,\frac{\,M^{\,\prime}}{M}\,\left(\frac{R}{a}\right)^{\textstyle{^{5}}}\,
 \frac{k_2}{Q}
 ~\\
 \label{57}\\
 \nonumber
 & - & \frac{21}{2}\,n\,e\,\frac{M}{\,M^{\,\prime}}\,\left(\frac{R^{\,\prime}}{a}\right)^{\textstyle{^{5}}}\,
 \frac{k_2^{\prime}}{Q^{\,\prime}}
 \;+\;O(e^2)\;+\,O(i^2)\;+\,O({i^{\,\prime\;}}^2)~~_{\textstyle{_{\textstyle ,}}}\qquad
 \ea
 \es
 the tidal parameters' values taken at the frequency $n$.

 To evaluate the circularisation timescale $\tau_e$, we need to know which term in the above expression is
 leading. For $M^{\,\prime}/M = 0.03$, the values of the factors  $\frac{\textstyle M}{\textstyle \,M^{\,\prime}}\,\left(\frac{\textstyle R^{\,\prime}}{\textstyle a}\right)^{\textstyle{^{5}}}$  and $\frac{\textstyle M^{\,\prime}}{\textstyle \,M}\,\left(\frac{\textstyle R}{\textstyle a}\right)^{\textstyle{^{5}}}$
  are comparable. Assuming that the planet is cooling down slower than the moon,
 i.e., that $k_2/Q\s>\s k_2^{\prime}/Q^{\,\prime}$, we conclude that the first term in expression (\ref{57}) is dominant. Accordingly,
 \ba
 \tau_e\s=\;\frac{2}{21}\;\frac{Q}{k_2}\;\frac{M}{M^{\,\prime}}\;\frac{1}{n}\s\left(\frac{a}{R}  \right)^5\;\,.
 \label{e}
 \ea
 Comparing this with the expression (\ref{p}) for planet despinning timescale $\tau_p$, we find that
 \ba
 \frac{\tau_e}{\tau_p}\,=\,\frac{2}{21}\,\frac{M^{\,\prime}}{M} \left(\frac{a}{R}\right)^{\textstyle{^{2}}} \approx 3\times 10^{-3}  \left(\frac{a}{R}\right)^{\textstyle{^{2}}}\,\;.
 \label{}
 \ea
 Although in realistic situations the right-hand side of this formula assumes values between $0.1$ and $0.3$, this in no way implies that the circularisation of orbit is attained before the planet gets despun. Recall that expression (\ref{e}) becomes valid only {\it{after}} the planet's synchronisation. Prior to that, a competition had been taking place: while the tides in the already synchronised moon were working to reduce the value of $e$, the tides in the still nonsynchronous planet were boosting $e$.
 So the circularisation process cannot even begin before the synchronism is reached.

 After the synchronism is attained, the subsequent circularisation never becomes complete, for two reasons. One is the influence of the triaxial figure of Mars on the moon's orbit. This influence is averaged out if the synchronism is achieved far from such a resonance, but becomes a great booster of the eccentricity if the synchronous orbit is resonant or near-resonant.  E.g. for Phobos, the eccentricity jumps $\Delta e$ due to the 2:1 and 3:1 resonances with Mars' figure, at $a_{\rm{_{Phobos}}} = 3.8\s R$ and $a_{\rm{_{Phobos}}} = 2.9\s R\s$, were 0.032 and $0.002\s$, correspondingly \citep{phobosdeimos, Yoder82}.

 The other reason is the Sun's gravity pull. The role of the Sun would have been especially strong, had the moon been formed or captured below the 2:1 MMR with the Sun at $a = 3.2R$. The moon's recession through this MMR would have produced a jump $\Delta e = 0.0085\s$ \citep{phobosdeimos, Yoder82}.

 [\s Mind misprints in the upper sentence on page 2 of \citet{phobosdeimos}. The words \,``{\it{...and a 1:1 resonance with the Sun at $a = 2.6 R_{\rm{_{Mars}}}$ when its pericentre rate equals the Martian mean motion}}''\, must be changed to:
       \,``{\it{...and a 2:1 resonance with the Sun at $a = 3.2 R_{\rm{_{Mars}}}$ when its pericentre rate equals one half of the Martian mean motion.}}'' Accordingly, in the third sentence from top, $\Delta e^{\rm{^{Sun}}}_{\rm{_{1:1}}} = 0.0085$ must be changed to $\Delta e^{\rm{^{Sun}}}_{\rm{_{2:1}}} = 0.0085\s$.\s]

 \section{Could Mars have ever stayed in a higher spin-orbit resonance with Nerio?\label{synchronism}}
  \setcounter{figure}{0}
 \renewcommand\thefigure{F\arabic{figure}}

  While a comprehensive analysis of Mars' synchronisation by the moon Nerio would amount to a separate project, we here explain in short why entrapment of a hot Mars into a higher spin-orbit resonance was impossible, even for a high initial eccentricity. Even if a not-yet-molten nascent Mars had somehow been trapped into a higher spin state,
  it was pushed out of that state towards synchronism after reaching the magma-ocean state.

  As mentioned in Section \ref{end-state}, a moon formed {\it in situ}, or produced by collision, or captured in the disk has a low initial eccentricity,
  wherefore neither of the partners end up in a higher-than-synchronous rotation state.

  %
  %
     \begin{figure}[H]
    \includegraphics[width=0.90\textwidth]{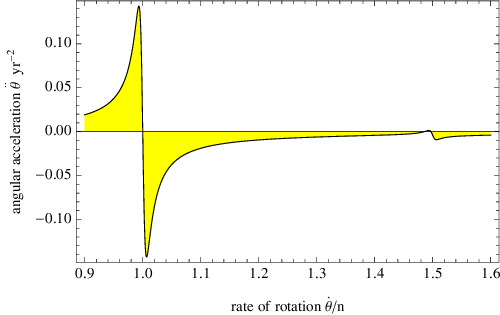}\\
             \caption{. \small The angular acceleration of a tidally perturbed rotator, as a function of its spin rate $\dot{\theta}$ divided by the mean motion $n\,$.
             The smaller term (vanishing in the 3:2 spin-orbit resonance) is superimposed on the main, semidiurnal term.  For sufficiently solid planets (i.e., for high mean viscosity values), the peaks are sharp, and the semidiurnal kink is squeezed toward the vertical axis. Its negative-valued right tail is, in the 3:2 resonance, overpowered by the small kink. The upper tip of the kink is residing above the horizontal axis. This creates a tidal trap: the planet is rotating in a stable equilibrium, because small deviations in  $\dot{\theta}/n$ render a torque that is always restoring the system back to the 3:2 spin state. For a hot planet (one of low mean viscosity), the peaks of the large kink spread broadly, the small kink ``sinks'', and the tidal trap is no more.
             \,(From \citeauthor{Noyelles} \citeyear{Noyelles})}
         \label{figure2}
     \end{figure}

  Assume that, despite this, the eccentricity somehow acquired a high initial value sufficient for Mars to get stuck in a higher spin state. To understand if this entrapment could happen, we resort to Figure \ref{figure2}.  Borrowed from \citet{Noyelles}, it depicts the dependence of the tidal torque (to be precise, of the angular acceleration) as a function of the spin rate $\dot{\theta}$ divided by the mean motion $n\s$. In the plot, a secondary term (one vanishing in the 3:2 spin-orbit resonance) is superimposed on the main, semidiurnal term.  The small kink overpowers the bias and produces a trap near the 3:2 resonance. In this trap, the rotator is spinning in a stable equilibrium, because small deviations in  $\dot{\theta}/n$ render a torque that is always restoring, a topic discussed at length in \citet{Noyelles}.

 It is important to spell out the exact reason why the magnitude of small kink can exceed the negative bias produced by the right tail of the semidiurnal kink. In the case of, e.g. Mercury, the reason is that Mercury is a solid planet. Had it been several times closer to the Sun, tidal heating would have put it into a semimolten state, with a much lower value of the mean viscosity of its mantle. The peaks of the semidiurnal torque would then have spread greatly (equation (\ref{wh}) in Appendix \ref{deta}), while the magnitudes of the involved terms would not have changed (equation (\ref{see}) in that appendix). The so widely spread big kink would have overpowered the small ones, and there would be no traps in higher-order spin-orbit resonances any more. In this situation, Mercury's capture into the 3:2 resonance would have been temporary; the planet would have been trapped, and would have left the resonance after getting tidally overheated. This issue was later addressed in detail by \citet{semiliquid}.

 Now suppose that owing to an unlikely but speculatively possible high value of the initial eccentricity, a nascent and not yet molten Mars went into
 a higher (say, 3:2) spin-orbit state with Nerio. At this point, the small kink in the plot was having its tip above the horizontal axis. As Mars was warming up and approaching the magma-ocean stage, its mean viscosity value was rapidly decreasing. Consequently, the peaks of the semidiurnal kink were spreading wide, leaving the small kinks no chance to keep the trap functioning. The small kink would lose the competition to the negative-valued tail of the main kink, and the rotator would leave the higher spin state, to continue its spin-down towards synchronism. {\it \`{A} propos}, tidal overheating is also the reason why close-in planets with appreciable eccentricities easily get caught into higher spin-orbit states~---~and then leave these states. This is what happened, e.g. with the planets $b$, $d$, and $e$ of TRAPPIST-1 \citep{trappist}.

 \bibliography{references}

\begin{thebibliography}{36}
\providecommand{\natexlab}[1]{#1}
\providecommand{\url}[1]{\texttt{#1}}
\expandafter\ifx\csname urlstyle\endcsname\relax
  \providecommand{\doi}[1]{doi: #1}\else
  \providecommand{\doi}{doi: \begingroup \urlstyle{rm}\Url}\fi

\bibitem[{Agnor} and {Hamilton}(2006)]{Agnor}
Craig~B. {Agnor} and Douglas~P. {Hamilton}.
\newblock {Neptune's capture of its moon Triton in a binary-planet
  gravitational encounter}.
\newblock \emph{Nature}, 441\penalty0 (7090):\penalty0 192--194, May 2006.
\newblock \doi{10.1038/nature04792}.

\bibitem[{Bagheri} et~al.(2021){Bagheri}, {Khan}, {Efroimsky}, {Kruglyakov},
  and {Giardini}]{phobosdeimos}
Amirhossein {Bagheri}, Amir {Khan}, Michael {Efroimsky}, Mikhail {Kruglyakov},
  and Domenico {Giardini}.
\newblock {Dynamical evidence for Phobos and Deimos as remnants of a disrupted
  common progenitor}.
\newblock \emph{Nature Astronomy}, 5:\penalty0 539--543, January 2021.
\newblock \doi{10.1038/s41550-021-01306-2}.

\bibitem[{Bagheri} et~al.(2022){Bagheri}, {Efroimsky}, {Castillo-Rogez},
  {Goossens}, {Plesa}, {Rambaux}, {Rhoden}, {Walterov{\'a}}, {Khan}, and
  {Giardini}]{Bagheri}
Amirhossein {Bagheri}, Michael {Efroimsky}, Julie {Castillo-Rogez}, Sander
  {Goossens}, Ana-Catalina {Plesa}, Nicolas {Rambaux}, Alyssa {Rhoden},
  Michaela {Walterov{\'a}}, Amir {Khan}, and Domenico {Giardini}.
\newblock {Tidal insights into rocky and icy bodies: an introduction and
  overview}.
\newblock \emph{Advances in Geophysics}, 63:\penalty0 231--320, 2022.
\newblock \doi{10.1016/bs.agph.2022.07.004}.

\bibitem[{Bou{\'e}} and {Efroimsky}(2019)]{BoueEfroimsky2019}
Gwena{\"e}l {Bou{\'e}} and Michael {Efroimsky}.
\newblock Tidal evolution of the keplerian elements.
\newblock \emph{Celestial Mechanics and Dynamical Astronomy}, 131:\penalty0 30,
  2019.
\newblock \doi{10.1007/s10569-019-9908-2}.

\bibitem[{Daradich} et~al.(2008){Daradich}, {Mitrovica}, {Matsuyama}, {Perron},
  {Manga}, and {Richards}]{matsuyama}
Amy {Daradich}, Jerry~X. {Mitrovica}, Isamu {Matsuyama}, J.~Taylor {Perron},
  Michael {Manga}, and Mark~A. {Richards}.
\newblock {Equilibrium rotational stability and figure of Mars}.
\newblock \emph{Icarus}, 194\penalty0 (2):\penalty0 463--475, 2008.
\newblock \doi{10.1016/j.icarus.2007.10.017}.

\bibitem[{Efroimsky}(2012)]{Efroimsky2012}
Michael {Efroimsky}.
\newblock {Bodily tides near spin-orbit resonances}.
\newblock \emph{Celestial Mechanics and Dynamical Astronomy}, 112\penalty0
  (3):\penalty0 283--330, March 2012.
\newblock \doi{10.1007/s10569-011-9397-4}.

\bibitem[{Efroimsky}(2015)]{Efroimsky2015}
Michael {Efroimsky}.
\newblock {Tidal Evolution of Asteroidal Binaries. Ruled by Viscosity. Ignorant
  of Rigidity.}
\newblock \emph{The Astronomical Journal}, 150\penalty0 (4):\penalty0 98, 2015.
\newblock \doi{10.1088/0004-6256/150/4/98}.

\bibitem[{Emelyanov}(2018)]{Emelyanov}
N.~V. {Emelyanov}.
\newblock {Influence of tides in viscoelastic bodies of planet and satellite on
  the satellite's orbital motion}.
\newblock \emph{Monthly Notices of the Royal Astronomical Society},
  479\penalty0 (1):\penalty0 1278--1286, 2018.
\newblock \doi{10.1093/mnras/sty1559}.

\bibitem[{Goldreich} and {Soter}(1966)]{soter}
Peter {Goldreich} and Steven {Soter}.
\newblock {Q in the Solar System}.
\newblock \emph{Icarus}, 5\penalty0 (1):\penalty0 375--389, 1966.
\newblock \doi{10.1016/0019-1035(66)90051-0}.

\bibitem[{Hiesinger} and {Head}(2004)]{syrtis}
H.~{Hiesinger} and J.~W. {Head}.
\newblock {The Syrtis Major volcanic province, Mars: Synthesis from Mars Global
  Surveyor data}.
\newblock \emph{Journal of Geophysical Research (Planets)}, 109\penalty0
  (E1):\penalty0 E01004, 2004.
\newblock \doi{10.1029/2003JE002143}.

\bibitem[Hunten(1979)]{Hunten}
Donald~M. Hunten.
\newblock Capture of phobos and deimos by photoatmospheric drag.
\newblock \emph{Icarus}, 37\penalty0 (1):\penalty0 113--123, 1979.
\newblock ISSN 0019-1035.
\newblock \doi{https://doi.org/10.1016/0019-1035(79)90119-2}.
\newblock URL
  \url{https://www.sciencedirect.com/science/article/pii/0019103579901192}.

\bibitem[{Hut}(1981)]{Hut}
P.~{Hut}.
\newblock {Tidal evolution in close binary systems}.
\newblock \emph{Astronomy and Astrophysics}, 99:\penalty0 126--140, 1981.

\bibitem[James et~al.(2004)James, Bagdassarov, M$\ddot{\mbox{u}}$ller, and
  Pinkerton]{James}
M.R James, N~Bagdassarov, K~M$\ddot{\mbox{u}}$ller, and H~Pinkerton.
\newblock Viscoelastic behaviour of basaltic lavas.
\newblock \emph{Journal of Volcanology and Geothermal Research}, 132\penalty0
  (2):\penalty0 99--113, 2004.
\newblock ISSN 0377-0273.
\newblock \doi{https://doi.org/10.1016/S0377-0273(03)00340-8}.
\newblock URL
  \url{https://www.sciencedirect.com/science/article/pii/S0377027303003408}.

\bibitem[{Kaula}(1964)]{Kaula}
William~M. {Kaula}.
\newblock {Tidal Dissipation by Solid Friction and the Resulting Orbital
  Evolution}.
\newblock \emph{Reviews of Geophysics and Space Physics}, 2:\penalty0 661--685,
  1964.
\newblock \doi{10.1029/RG002i004p00661}.

\bibitem[{Konopliv} et~al.(2011){Konopliv}, {Asmar}, {Folkner}, {Karatekin},
  {Nunes}, {Smrekar}, {Yoder}, and {Zuber}]{Konopliv_2011}
Alex~S. {Konopliv}, Sami~W. {Asmar}, William~M. {Folkner}, {\"O}zg{\"u}r
  {Karatekin}, Daniel~C. {Nunes}, Suzanne~E. {Smrekar}, Charles~F. {Yoder}, and
  Maria~T. {Zuber}.
\newblock {Mars high resolution gravity fields from MRO, Mars seasonal gravity,
  and other dynamical parameters}.
\newblock \emph{Icarus}, 211\penalty0 (1):\penalty0 401--428, January 2011.
\newblock \doi{10.1016/j.icarus.2010.10.004}.

\bibitem[{Konopliv} et~al.(2016){Konopliv}, {Park}, and
  {Folkner}]{Konopliv2016}
Alex~S. {Konopliv}, Ryan~S. {Park}, and William~M. {Folkner}.
\newblock {An improved JPL Mars gravity field and orientation from Mars orbiter
  and lander tracking data}.
\newblock \emph{Icarus}, 274:\penalty0 253--260, 2016.
\newblock \doi{10.1016/j.icarus.2016.02.052}.

\bibitem[{Konopliv} et~al.(2020){Konopliv}, {Park}, {Rivoldini}, {Baland}, {Le
  Maistre}, {Van Hoolst}, {Yseboodt}, and {Dehant}]{Konopliv}
Alex~S. {Konopliv}, Ryan~S. {Park}, Attilio {Rivoldini}, Rose-Marie {Baland},
  Sebastien {Le Maistre}, Tim {Van Hoolst}, Marie {Yseboodt}, and Veronique
  {Dehant}.
\newblock {Detection of the Chandler Wobble of Mars From Orbiting Spacecraft}.
\newblock \emph{Geophysical Research Letters}, 47\penalty0 (21):\penalty0
  e90568, 2020.
\newblock \doi{10.1029/2020GL090568}.

\bibitem[{Lainey} et~al.(2012){Lainey}, {Karatekin}, {Desmars}, {Charnoz},
  {Arlot}, {Emelyanov}, {Le Poncin-Lafitte}, {Mathis}, {Remus}, {Tobie}, and
  {Zahn}]{Lainey}
Val{\'e}ry {Lainey}, {\"O}zg{\"u}r {Karatekin}, Josselin {Desmars},
  S{\'e}bastien {Charnoz}, Jean-Eudes {Arlot}, Nicolai {Emelyanov}, Christophe
  {Le Poncin-Lafitte}, St{\'e}phane {Mathis}, Fran{\c{c}}oise {Remus}, Gabriel
  {Tobie}, and Jean-Paul {Zahn}.
\newblock {Strong Tidal Dissipation in Saturn and Constraints on Enceladus'
  Thermal State from Astrometry}.
\newblock \emph{The Astrophysical Journal}, 752\penalty0 (1):\penalty0 14,
  2012.
\newblock \doi{10.1088/0004-637X/752/1/14}.

\bibitem[{Leinhardt} et~al.(2012){Leinhardt}, {Ogilvie}, {Latter}, and
  {Kokubo}]{Leinhardt}
Z.~M. {Leinhardt}, G.~I. {Ogilvie}, H.~N. {Latter}, and E.~{Kokubo}.
\newblock {Tidal disruption of satellites and formation of narrow rings}.
\newblock \emph{Monthly Notices of the Royal Astronomical Society},
  424\penalty0 (2):\penalty0 1419--1431, 2012.
\newblock \doi{10.1111/j.1365-2966.2012.21328.x}.

\bibitem[{Makarov}(2015)]{semiliquid}
Valeri~V. {Makarov}.
\newblock {Equilibrium Rotation of Semiliquid Exoplanets and Satellites}.
\newblock \emph{The Astrophysical Journal}, 810\penalty0 (1):\penalty0 12,
  2015.
\newblock \doi{10.1088/0004-637X/810/1/12}.

\bibitem[{Makarov} and {Efroimsky}(2023)]{pathways}
Valeri~V. {Makarov} and Michael {Efroimsky}.
\newblock {Pathways of survival for exomoons and inner exoplanets}.
\newblock \emph{Astronomy and Astrophyscs}, 672:\penalty0 A78, 2023.
\newblock \doi{10.1051/0004-6361/202245533}.

\bibitem[{Makarov} et~al.(2018){Makarov}, {Berghea}, and {Efroimsky}]{trappist}
Valeri~V. {Makarov}, Ciprian~T. {Berghea}, and Michael {Efroimsky}.
\newblock {Spin-orbital Tidal Dynamics and Tidal Heating in the TRAPPIST-1
  Multiplanet System}.
\newblock \emph{The Astrophysical Journal}, 857\penalty0 (2):\penalty0 142,
  2018.
\newblock \doi{10.3847/1538-4357/aab845}.

\bibitem[{Nesvorn{\'y}} et~al.(2023){Nesvorn{\'y}}, {Roig}, {Vokrouhlick{\'y}},
  {Bottke}, {Marchi}, {Morbidelli}, and {Deienno}]{Morbidelli}
D.~{Nesvorn{\'y}}, F.~V. {Roig}, D.~{Vokrouhlick{\'y}}, W.~F. {Bottke},
  S.~{Marchi}, A.~{Morbidelli}, and R.~{Deienno}.
\newblock {Early bombardment of the moon: Connecting the lunar crater record to
  the terrestrial planet formation}.
\newblock \emph{Icarus}, 399:\penalty0 115545, 2023.
\newblock \doi{10.1016/j.icarus.2023.115545}.

\bibitem[{Noyelles} et~al.(2014){Noyelles}, {Frouard}, {Makarov}, and
  {Efroimsky}]{Noyelles}
Beno{\^\i}t {Noyelles}, Julien {Frouard}, Valeri~V. {Makarov}, and Michael
  {Efroimsky}.
\newblock {Spin-orbit evolution of Mercury revisited}.
\newblock \emph{Icarus}, 241:\penalty0 26--44, October 2014.
\newblock \doi{10.1016/j.icarus.2014.05.045}.

\bibitem[{Quillen} et~al.(2019){Quillen}, {Martini}, and {Nakajima}]{Quillen}
Alice~C. {Quillen}, Larkin {Martini}, and Miki {Nakajima}.
\newblock {Near/far side asymmetry in the tidally heated Moon}.
\newblock \emph{Icarus}, 329:\penalty0 182--196, 2019.
\newblock \doi{10.1016/j.icarus.2019.04.010}.

\bibitem[{Rufu} et~al.(2017){Rufu}, {Aharonson}, and {Perets}]{rufu}
R.~{Rufu}, O.~{Aharonson}, and H.~B. {Perets}.
\newblock {A multiple-impact origin for the Moon}.
\newblock \emph{Nature Geoscience}, 10\penalty0 (2):\penalty0 89--94, 2017.
\newblock \doi{10.1038/ngeo2866}.

\bibitem[Ruiz-Bonilla et~al.(2020)Ruiz-Bonilla, Eke, Kegerreis, Massey, and
  Teodoro]{Ruiz}
S~Ruiz-Bonilla, V~R Eke, J~A Kegerreis, R~J Massey, and L~F~A Teodoro.
\newblock {The effect of pre-impact spin on the Moon-forming collision}.
\newblock \emph{Monthly Notices of the Royal Astronomical Society},
  500\penalty0 (3):\penalty0 2861--2870, 12 2020.
\newblock ISSN 0035-8711.
\newblock \doi{10.1093/mnras/staa3385}.
\newblock URL \url{https://doi.org/10.1093/mnras/staa3385}.

\bibitem[{Samuel} et~al.(2019){Samuel}, {Lognonn{\'e}}, {Panning}, and
  {Lainey}]{samuel}
H.~{Samuel}, P.~{Lognonn{\'e}}, M.~{Panning}, and V.~{Lainey}.
\newblock {The rheology and thermal history of Mars revealed by the orbital
  evolution of Phobos}.
\newblock \emph{Nature}, 569\penalty0 (7757):\penalty0 523--527, 2019.
\newblock \doi{10.1038/s41586-019-1202-7}.

\bibitem[Segatz et~al.(1988)Segatz, Spohn, Ross, and Schubert]{segatz88}
M.~Segatz, T.~Spohn, M.~N. Ross, and G.~Schubert.
\newblock {T}idal {D}issipation, {S}urface {H}eat {F}low, and {F}igure of
  {V}iscoelastic {M}odels of {Io}.
\newblock \emph{Icarus}, 75\penalty0 (2):\penalty0 187--206, 1988.
\newblock \doi{10.1016/0019-1035(88)90001-2}.

\bibitem[{Seidelmann} and {Urban}(2013)]{Seidelmann}
P.~Kenneth {Seidelmann} and S.~E. {Urban}.
\newblock \emph{{Explanatory Supplement to the Astronomical Almanac, Third
  Edition}}.
\newblock 2013.
\newblock ISBN 978-1-891389-85-6.

\bibitem[{Smith} and {Zuber}(1996)]{smith}
D.~E. {Smith} and M.~T. {Zuber}.
\newblock {The Shape of Mars and the Topographic Signature of the Hemispheric
  Dichotomy}.
\newblock In \emph{Lunar and Planetary Science Conference Series}, volume~27,
  pages 1221--1222, March 1996.

\bibitem[Tiesinga et~al.(2021)Tiesinga, Mohr, Newell, and
  Taylor]{gravitational_constant}
Eite Tiesinga, Peter~J. Mohr, David~B. Newell, and Barry~N. Taylor.
\newblock {CODATA} recommended values of the fundamental physical constants:
  2018.
\newblock \emph{Reviews of Modern Physics}, 93:\penalty0 025010, Jun 2021.
\newblock \doi{10.1103/RevModPhys.93.025010}.
\newblock URL \url{https://link.aps.org/doi/10.1103/RevModPhys.93.025010}.

\bibitem[{Walterov{\'a}} et~al.(2023){Walterov{\'a}}, {B{\v{e}}hounkov{\'a}},
  and {Efroimsky}]{walterova}
Michaela {Walterov{\'a}}, Marie {B{\v{e}}hounkov{\'a}}, and Michael
  {Efroimsky}.
\newblock {Is There a Semi-Molten Layer at the Base of the Lunar Mantle?}
\newblock \emph{Journal of Geophysical Research (Planets)}, 128\penalty0
  (7):\penalty0 e2022JE007652, 2023.
\newblock \doi{10.1029/2022JE007652}.

\bibitem[Williams and Zugger(2024)]{Williams_and_Zugger}
Darren~M. Williams and Michael~E. Zugger.
\newblock Forming massive terrestrial satellites through binary-exchange
  capture.
\newblock \emph{The Planetary Science Journal}, 5\penalty0 (9):\penalty0 208,
  2024.
\newblock \doi{10.3847/PSJ/ad5a9a}.
\newblock URL \url{https://dx.doi.org/10.3847/PSJ/ad5a9a}.

\bibitem[{Yoder}(1982)]{Yoder82}
C.~F. {Yoder}.
\newblock {Tidal rigidity of Phobos}.
\newblock \emph{Icarus}, 49:\penalty0 327--346, 1982.
\newblock \doi{10.1016/0019-1035(82)90040-9}.

\bibitem[{Zuber} and {Smith}(1997)]{zuber}
M.~T. {Zuber} and D.~E. {Smith}.
\newblock {Mars without Tharsis}.
\newblock \emph{Journal of Geophysical Research (Planets)}, 102\penalty0
  (E12):\penalty0 28673--28686, 1997.
\newblock \doi{10.1029/97JE02527}.

\end{thebibliography}
 \bibliographystyle{plainnat}

 \enlargethispage{\baselineskip}
  \enlargethispage{\baselineskip}
   \enlargethispage{\baselineskip}

 \end{document}